\documentclass[iop]{emulateapj}
\usepackage{color}
\usepackage{graphicx}
\usepackage{longtable}
\usepackage{lineno}
\usepackage{draftcopy}
\usepackage{threeparttable}
\usepackage{multirow}
\usepackage{rotating}
\slugcomment{version \today: fm}
\shorttitle{}
\shortauthors{}

\begin{document}
\title{The WISE Blazar-like Radio-Loud Sources:\\ 
	an All-Sky Catalog of Candidate $\gamma$-ray Blazars}
\author{R.~D'Abrusco\altaffilmark{1}, F.~Massaro\altaffilmark{2}, A.~Paggi\altaffilmark{1}, 
	    H.~A.~Smith\altaffilmark{1}, N.~Masetti\altaffilmark{3}, M.~Landoni\altaffilmark{4} 
	    \& G.~Tosti\altaffilmark{5,6}}

\altaffiltext{1}{Harvard - Smithsonian Center for Astrophysics, 60 Garden Street, Cambridge, 
MA 02138, USA}
\altaffiltext{2}{Yale Center for Astronomy and Astrophysics, Physics Department, Yale University, 
PO Box 208120, New Haven, CT 06520-8120, USA}
\altaffiltext{3}{INAF/IASF di Bologna, via Gobetti 101, I-40129 Bologna, Italy}
\altaffiltext{4}{INAF/Osservatorio Astronomico di Brera, Via E. Bianchi 46, 23807
Merate, Italy}
\altaffiltext{5}{Dipartimento di Fisica, Universit\`a degli Studi di Perugia, 06123 Perugia, Italy}
\altaffiltext{6}{Istituto Nazionale di Fisica Nucleare, Sezione di Perugia, 06123 Perugia, Italy}

\begin{abstract}

We present a catalog of radio-loud candidate $\gamma$-ray emitting blazars with WISE mid-infrared 
colors similar to the colors of confirmed $\gamma$-ray blazars. The catalog is assembled from WISE 
sources detected in all four WISE filters, with colors compatible with the three-dimensional {\it locus} of 
the WISE $\gamma$-ray emitting blazars, and which can be spatially cross-matched with radio sources 
from either one 
of the three radio surveys: NVSS, FIRST and/or SUMSS. Our initial WISE selection uses a slightly modified 
version of previously successful algorithms. We then select only the radio-loud sources using a measure of 
the radio-to-IR flux, the $q_{22}$ parameter, which is analogous to the $q_{24}$ parameter known in the 
literature but which instead uses the WISE band-four flux at 22 $\mu$m. Our final catalog contains 
7855 sources classified as BL Lacs, FSRQs or mixed candidate blazars; 1295 of these sources 
can be spatially re-associated with confirmed blazars.  
We describe the properties of the final catalog of WISE blazar-like radio-loud sources and consider possible 
contaminants.  Finally, we discuss why this large catalog of candidate $\gamma$-ray emitting blazars 
represents a new and useful resource to address the problem of finding low energy counterparts to 
currently unidentified high-energy sources.

\end{abstract}


\section{Introduction}
\label{sec:intro}

The largest known class of $\gamma$-ray sources is represented by the 
rarest class of Active Galactic Nuclei (AGNs), blazars~\citep[e.g.][]{abdo2010,nolan2012}.
This population of radio-loud sources is mainly characterized by flat radio spectra and superluminal 
motions, variable and high polarization from the radio to the optical band~\citep{urry1995,massaro2009}.
Their emission is strongly dominated by non-thermal radiation over the entire electromagnetic spectrum,
featuring two broad components in their spectral energy distributions: the low-energy one peaking 
between the IR and the X-ray and the high-energy one exhibiting its maximum in the $\gamma$-ray
energies.

Blazars are historically divided in two main classes on the basis of their optical spectra.
The first class includes the BL Lac objects, characterized by featureless spectra with emission/absorption 
lines of equivalent width lower than 5\AA~\citep{stickel1991,stocke1991,laurent1999}. 
The second class is represented by the flat spectrum radio quasars (FSRQs), that show normal quasar-like 
spectra. In the following we adopt the nomenclature proposed in the Multi-wavelength Blazar 
Catalog\footnote{http://www.asdc.asi.it/bzcat/}~\citep[BZCat,][]{massaro2009,massaro2011}, that labels  
BL Lac objects as BZBs and FSRQs as BZQs.

The ROMA-BZCat is based on by-eye inspection of multi-frequency data and the extensive review 
of the literature for each member. The minimal requirements that a source has to meet to be included in the BZCat, 
are the optical identification and/or availability of optical spectrum, X-ray luminosity 
equal or larger than 10$^{43}$ erg $s^{-1}$ and a compact radio morphology. These stringent requirements and 
the use of heterogeneous data make for a very reliable but incomplete list of {\it bona fide} 
blazars. The selection of large, homogeneous samples of blazars is intrinsically difficult, because of 
their peculiar spectral characteristics and extreme variability. New simpler criteria can be useful to extract 
larger and less incomplete samples of candidate blazars, whose nature has to be confirmed through additional 
follow-up observations.

We have recently discovered that $\gamma$-ray emitting blazars have infrared colors that 
distinguish them from other galactic and extragalactic sources in the three-dimensional colors 
space of the mid-IR WISE magnitudes~\citep{massarof2011,dabrusco2012}. 
We used this result to devise a new method for the association of the {\it Fermi} Large
Area Telescope (LAT) unidentified $\gamma$-ray sources through a parametrization of the 
region occupied by $\gamma$-ray blazars in the WISE colors space, the so-called WISE blazars
{\it locus}~\citep{dabrusco2013,massaro2013a}. 

In this paper, we present a catalog of candidate $\gamma$-ray emitting blazars extracted 
from the AllWISE Data Release~\footnote{http://wise2.ipac.caltech.edu/docs/release/allwise/expsup/}. 
This catalog is composed of radio-loud WISE sources detected in all four WISE filters, whose 
mid-IR colors are similar to the typical colors of confirmed $\gamma$-ray emitting 
blazars~\citep[see][]{dabrusco2013}, spatially associated to a radio source and selected as radio-loud. 
Hereinafter, such sources will be called ``WISE Blazar-like RAdio-Loud Sources'', or WIBRaLS.

The paper is organized as follows: in Section~\ref{sec:selection} we give a brief 
summary of the procedure used to select the WIBRaLS and the basic information about
the final catalog. Specifically, in Section~\ref{subsec:wise_selection} we discuss 
the method used to select the WISE sources with IR colors similar to the colors of the $\gamma$-ray 
emitting blazars. Section~\ref{subsec:spatial_selection} is devoted to the description of 
the technique used to perform the spatial association of the WISE sources with the radio counterparts, 
and in Section~\ref{subsec:q22_selection} we 
introduce the radio-loudness parameter $q_{22}$ and discuss its application to select radio-loud 
sources among the sample of WISE blazar-like sources with a radio association.
In Section~\ref{sec:catalog}, we describe the final catalog of WIBRaLS, and in Section~\ref{sec:discussion}
we compare the WIBRaLS with other WISE-based techniques optimized 
for the selection of AGNs (Section~\ref{subsec:comparison}) and with the VERONCAT~\citep{veron2010} 
(Section~\ref{subsec:veroncat}), to indirectly characterize the nature of the sources 
in our catalog and assess possible contamination from non-blazars. Finally, 
in Section~\ref{sec:summary} we summarize the results and draw our 
conclusions. We use cgs units and spectral indices are based on the definition of 
the flux density as $S_{\nu}\!=\!\nu^{-\alpha}$.

The WISE magnitudes in the $[3.4]$, $[4.6]$, $[12]$, $[22]\mu$m nominal filters 
are in the Vega system. The values of three WISE magnitudes, namely $[3.4]$, $[4.6]$ 
and $[12]$, and of the colors derived using those magnitudes, have been 
corrected for galactic extinction according to the extinction law presented by~\cite{draine2003}. 

\begin{figure}[] 
	\includegraphics[height=9cm,width=8.5cm,angle=0]{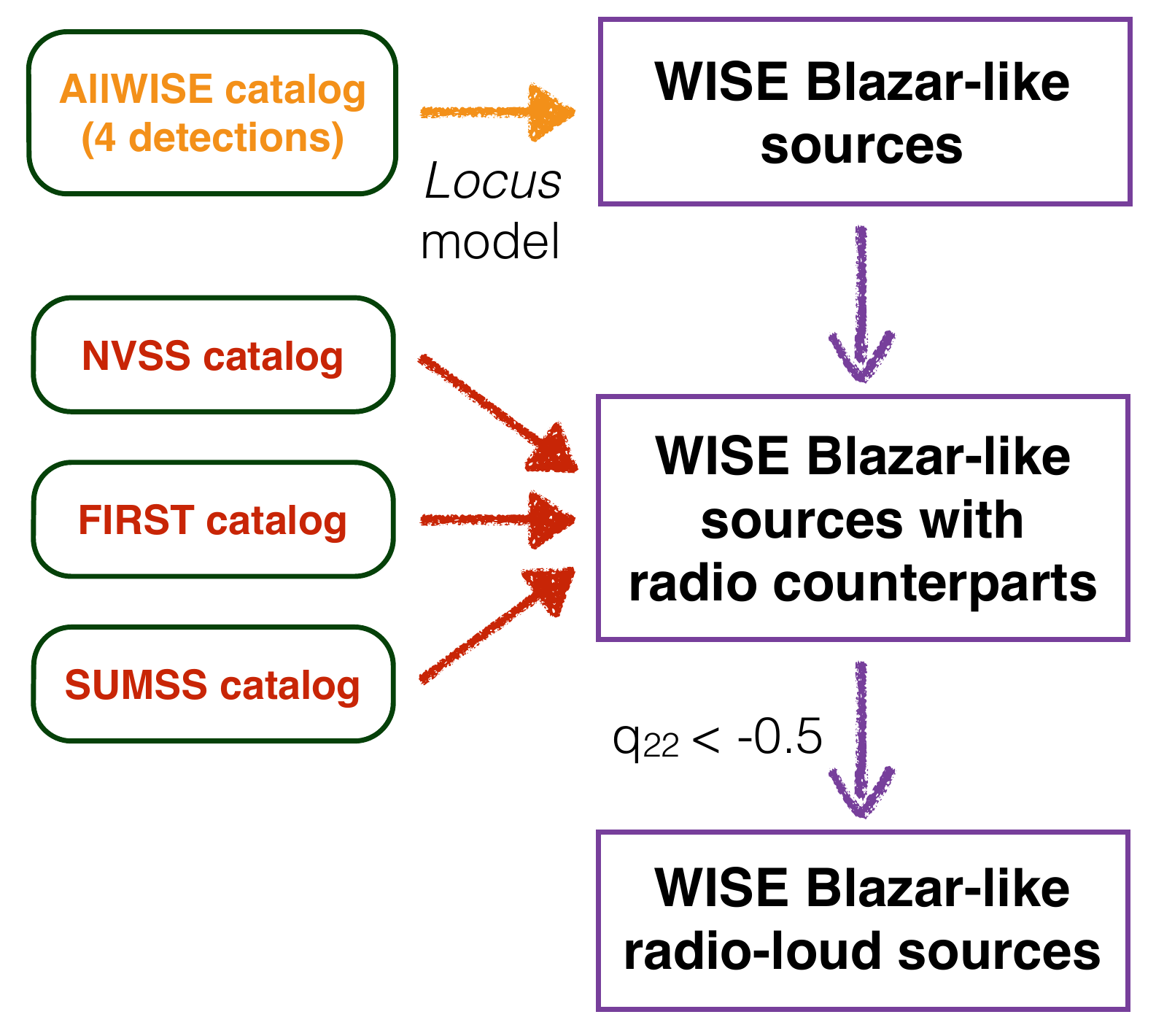}	
      	\caption{Schematic representation of the workflow for the extraction of the WIBRaLS.}
       	\label{fig:workflow}
\end{figure}

\section{The Selection of the WISE Blazar-like Radio-Loud Sources}
\label{sec:selection}

The three steps followed to select WISE blazar-like radio-loud sources can be summarized as follows:

\begin{itemize}
	\item WISE sources detected in all four filters $[3.4]$, $[4.6]$, $[12]$, $[22]\mu$m 
	are selected according to their mid-IR colors, using a slightly modified 
	version of the technique for the association of high-energy sources with WISE candidate blazars 
	presented by~\cite{dabrusco2013} (see Section~\ref{subsec:wise_selection}). We designate 
	these sources as ``WISE blazar-like sources''. 
	\item The WISE blazar-like sources selected with the method described in 
	Section~\ref{subsec:wise_selection} are positionally cross-matched with the catalogs of sources
	detected in three
	different radio surveys: NRAO VLA Sky Survey - 
	NVSS~\citep{condon1998}\footnote{http://heasarc.gsfc.nasa.gov/W3Browse/all/nvss.html}, 
	the Sydney University Molonglo Sky Survey - 
	SUMSS~\citep{mauch2003}\footnote{http://heasarc.gsfc.nasa.gov/W3Browse/all/sumss.html} and 
	the Faint Images of the Radio Sky at Twenty-cm survey - 
	FIRST~\citep{becker1995}\footnote{http://heasarc.gsfc.nasa.gov/W3Browse/all/first.html}. Only  
	the WISE blazar-like sources that can be associated to 
	at least one radio source from either one of three radio surveys within the maximum 
	radial distances discussed in Section~\ref{subsec:spatial_selection} are further 
	considered. 
	\item We retain only the WISE blazar-like 
	sources with radio counterpart that satisfy the radio-loudness criterion $q_{22}\!\leq\!-0.5$
	(Section~\ref{subsec:q22_selection}), in order to minimize the contamination in our catalog 
	from sources whose radio emission is not associated to AGN activity.
\end{itemize}

We anticipate that the final number of unique WIBRaLS selected with our method is 
7855. A workflow representing the procedure for the extraction 
of the catalog of WIBRaLS is shown in Figure~\ref{fig:workflow}. In the following sections, 
we will discuss in details the three steps summarized above.

\subsection{Extraction of the WISE blazar-like sources}
\label{subsec:wise_selection}

The method for the extraction of the WISE blazar-like sources used in this paper is based on the 
technique for the association of the unidentified $\gamma$-ray sources presented 
by~\cite{dabrusco2013}. Here a modified version of that method will be used to extract the WISE sources with 
IR colors similar to the typical colors of the $\gamma$-ray emitting blazars from the whole sky, 
without any spatial constraint, while the association method in~\cite{dabrusco2013} selected 
candidate blazars located in the regions of the sky where $\gamma$-ray source are detected.

In what follows, a brief summary of the association method will be given~\citep[see][for a detailed 
description of the dataset and method]{dabrusco2013}.~\cite{dabrusco2012} found that $\gamma$-ray 
emitting blazars occupy a narrow region of the three-dimensional color space associated to the four 
WISE filters, called the {\it locus}. In~\cite{dabrusco2013}, the {\it locus} was defined using a 
sample of confirmed {\it Fermi} $\gamma$-ray blazars, based on the ROMA-BZCat~\citep{massaro2011} 
and the {\it Fermi} LAT 2FGL~\citep{nolan2012}, associated to WISE counterparts 
detected in all four WISE filters ($[3.4]$, $[4.6]$, $[12]$, $[22]\mu$m). The 
{\it locus} has been modeled in the three-dimensional space generated by the 
Principal Components (PCs) of the IR colors distribution of the WISE sources of the 
{\it locus} itself. 

This model is composed of a set of three coaxial cylindrical regions aligned to the PC1 
axis. Two cylinders are dominated by blazars of the same spectral classes, namely BZB and
BZQ, while the third cylinder is positioned between these two cylinders and
contains a mixed population of both BZBs and BZQs. The ranges of PC1 coordinates spanning the heights 
of the three cylinders were determined so that the two extreme cylinders contain at least 75\% of {\it locus} 
sources classified as BZBs and BZQs, respectively. The height of the Mixed cylinder was set accordingly. 
The radii of the three cylinders are determined so that 90\% of 
the sources whose PC1 coordinate lies within the PC1 ranges of each cylinder, lie within the respective 
cylinder (see Figure 3 and Table 2 of~\citealt{dabrusco2013}). 

The position of a generic WISE source relative to each of the three distinct
cylinders in the model of the {\it locus}, is used to choose the sources most likely to 
have the typical WISE colors of the $\gamma$-ray 
emitting blazars. We called the quantitative measure of the compatibility of the position of a 
generic WISE source with each of the three cylinders of the model, separately, the score. 
The score can be calculated in two steps: 
1) the WISE colors of the source (and corresponding uncertainties) 
are projected into the PC space, where an error ellipsoid is defined; 2) the position of the 
ellipsoid relative to the {\it locus} model is translated to a numeric value, the {\it score}, which varies 
continuously between zero and one, and is weighted by the volume of the error ellipsoid in the PC 
space~\citep[see Section 3.2 of][for the definition of {\it score}]{dabrusco2013}. 
In general, the larger the {\it score values}, the closer the source to the {\it locus} model and 
the more similar the WISE colors to the colors of the confirmed $\gamma$-ray emitting blazars. 

In~\cite{dabrusco2013} the sources with null scores were discarded, while WISE sources with non-null 
{\it scores} were classified in three classes, namely A, B and C, based on {\it score} thresholds 
defined as the 90\%-th, 60\%-th and 20\%-th percentiles of the {\it score} distribution of the {\it locus} 
sample, separately for each of the three regions of the {\it locus}. The three classes A, B and C are sorted 
according to decreasing compatibility with the model of the \emph{locus}: class A sources are considered to be 
the most likely WISE blazar-like sources, while the positions of class B and class C sources are still 
compatible with the model of the \emph{locus} but at a lesser degree than class A sources. 

\begin{figure}[] 
	\includegraphics[height=9cm,width=8.5cm,angle=0]{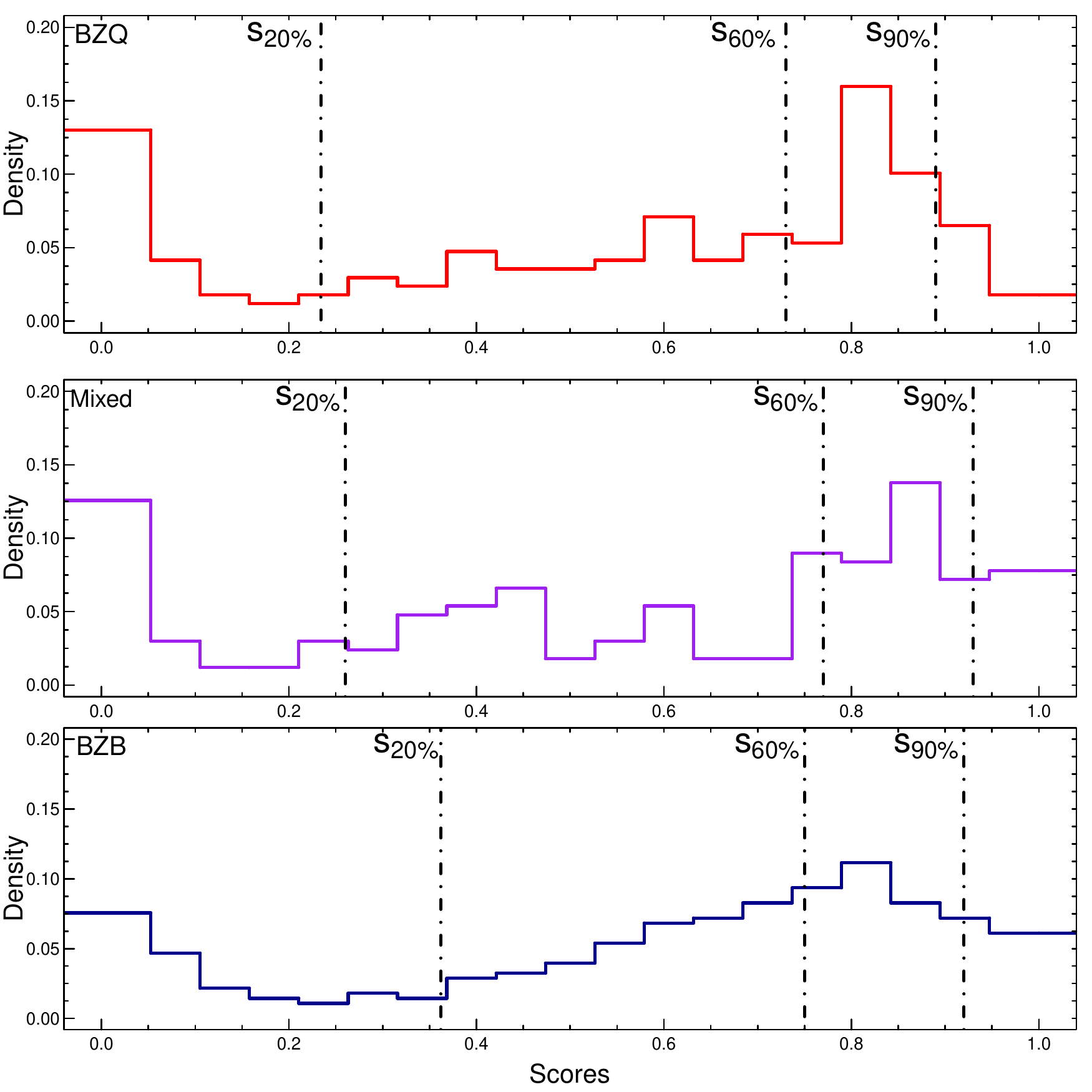}	
      	\caption{Histograms of distributions of {\it score} values for the sources in the sample 
       	used to define the {\it locus} model (see Section~\ref{subsec:wise_selection})
	for the three regions of the \emph{locus} model occupied by BZQs, the BZBs and
        	in the mixed region (upper, lower and mid panels respectively). The three 
       	dashed vertical lines in each panel represent the values of 
       	the {\it score} associated with the 20\%-th, 60\%-th and 90\%-th
       	percentiles for BZBs, BZQs and mixed sources respectively.}
       	\label{fig:scoreswgs}
\end{figure}

In this paper, we use the same approach
described in~\cite{dabrusco2013} to determine the value of the {\it score} thresholds, except for the 
percentile associated to the lowest threshold. In order to select the largest possible number of WISE
blazar-like sources, we set the three thresholds to the 20\%-th, 60\%-th and 90\%-th percentiles of the 
{\it score} distribution of the {\it locus} sample (see Figure~\ref{fig:scoreswgs} and compare with Figure 6 
in~\citealt{dabrusco2013}). The adoption of a lower threshold for the definition of the class C
sources makes our selection more complete but, at the same time, potentially increases the 
contamination of the class C from sources that only marginally have WISE colors similar to the colors
of the confirmed WISE $\gamma$-ray emitting blazars. The presence of class C contaminants is 
mitigated by considering only the WISE blazar-like sources which can be positionally associated 
with sources in either one of the three radio surveys (NVSS, FIRST, SUMSS) used in this 
paper (Section~\ref{subsec:spatial_selection}). 

For each distinct region of the model (BZB, BZQ or Mixed), class A sources have score $s\!\geq\!s_{90\%}$, 
class B sources have score $s_{60\%}\!\leq\!s\!<\!s_{90\%}$ and class C sources have score 
$s_{20\%}\!\leq\!s\!<\!s_{60\%}$. The values of the {\it score} thresholds derived from the score 
distributions of the sources in the {\it locus} for the three regions of the model are reported in 
Table~\ref{tab:thresholds}. Each source with score larger than the $s_{20\%}$ threshold for one of the cylinders 
of the model, is assigned the corresponding type (BZB, BZQ or Mixed). The WISE blazar-like sources whose type
are BZB or BZQ have IR colors similar to the colors of the {\it bona fide} WISE-detected 
$\gamma$-ray emitting blazars classified as BZB or BZQ, respectively. The type Mixed, conversely, 
does not indicate any particular spectral class since the Mixed cylinder contains comparable fractions
of both BZBs and BZQs.
 
The total number of WISE blazar-like sources selected by our method is 265170 (see Table~\ref{tab:samples}), 
split in 32789 BZB-type candidates, 169703 BZQ-type candidates and 62678 sources compatible with the Mixed
region of the {\it locus}. The sample of 265170 WISE blazar-like sources is split in 3554 
ranked as Class A, 17500 as Class B and the remaining 244116 classified as Class C candidates.
The Class C candidates represent $\sim\!92\%$ of the total number of WISE blazar-like sources, while
the Class A and Class B candidates account only for the $\sim\!1.3\%$ and $\sim\!6.6\%$, respectively. 
These fractions result from the conservative choices of the {\it score} thresholds used to define the 
classes of the sources based on their WISE colors and uncertainties (see~\citealt{dabrusco2013} for
more details). 

\begin{table}
	\begin{center}
	\caption{Values of the {\it score} thresholds $s_{20\%}$, $s_{60\%}$ and $s_{90\%}$, used for 
	the extraction of the WISE blazar-like sources. These values are determined as the 
	20\%-th, 60\%-th and 90\%-th percentiles of the distribution of {\it scores} of the {\it locus} 
	sample split by BZB, Mixed and BZB regions.}
	\begin{tabular}{cccc}
	\tableline
	\tableline
					& BZB 	& Mixed	& BZQ  	\\
	\tableline
	$s_{20\%}$		& 0.36	& 0.26	& 0.23	\\
	$s_{60\%}$		& 0.75	& 0.77	& 0.79	\\
	$s_{90\%}$		& 0.92	& 0.93	& 0.89	\\	
	\tableline
	\label{tab:thresholds}
	\end{tabular}
	\end{center}
\end{table}

\subsection{Spatial crossmatch with radio catalogs}
\label{subsec:spatial_selection}

In order to determine the optimal radius for the spatial association of the WISE 
blazar-like sources with the radio sources in the NVSS, the SUMSS and FIRST catalogs, 
we used a modified version of the procedure used by~\cite{donoso2009} and~\cite{best2005}. 
In these two papers, the authors determined the optimal radius for the spatial association of NVSS 
and FIRST radio sources to optical sources in SDSS by setting a threshold on the fraction of spurious 
associations (i.e. the contamination) obtained for different values of the maximum association radius.

In this paper, we will use as optimal association radius the radial distance corresponding to a given fixed
efficiency of the selection $e_{\mathrm{thr}}\!=\!99\%$\footnote{In this paper, we call ``efficiency'' the same 
quantity called ``reliability'' by~\cite{best2005}}, corresponding to a contamination $c_{\mathrm{thr}}\!=\!1\%$
where the contamination is defined as $c(r)\!=\!100\%\!-\!e(r)$. This choice of the efficiency has the result 
of minimizing the fraction of spurious sources 
selected and optimize the success likelihood of the follow-up observations required to confirm their 
nature, at the cost of a limited completeness.
The efficiency of the selection is defined as the number of sources 
around real radio positions $n^{\mathrm{real}}(r)$ minus the number of sources around the mock radio 
positions $n^{\mathrm{mock}}(r)$, expressed as a fraction of the number of real cross-matches. For a given 
radius $r$, the efficiency $e(r)$ is defined as:

\begin{equation}
	e(r)\!=\!100\!*\!\Big(\frac{n^{\mathrm{real}}(r)\!-\!n^{\mathrm{mock}}(r)}{n^{\mathrm{real}}(r)}\Big)
	\label{eq:contamination}
\end{equation}

\begin{figure*}[]
	\includegraphics[height=7cm,width=6cm,angle=0]{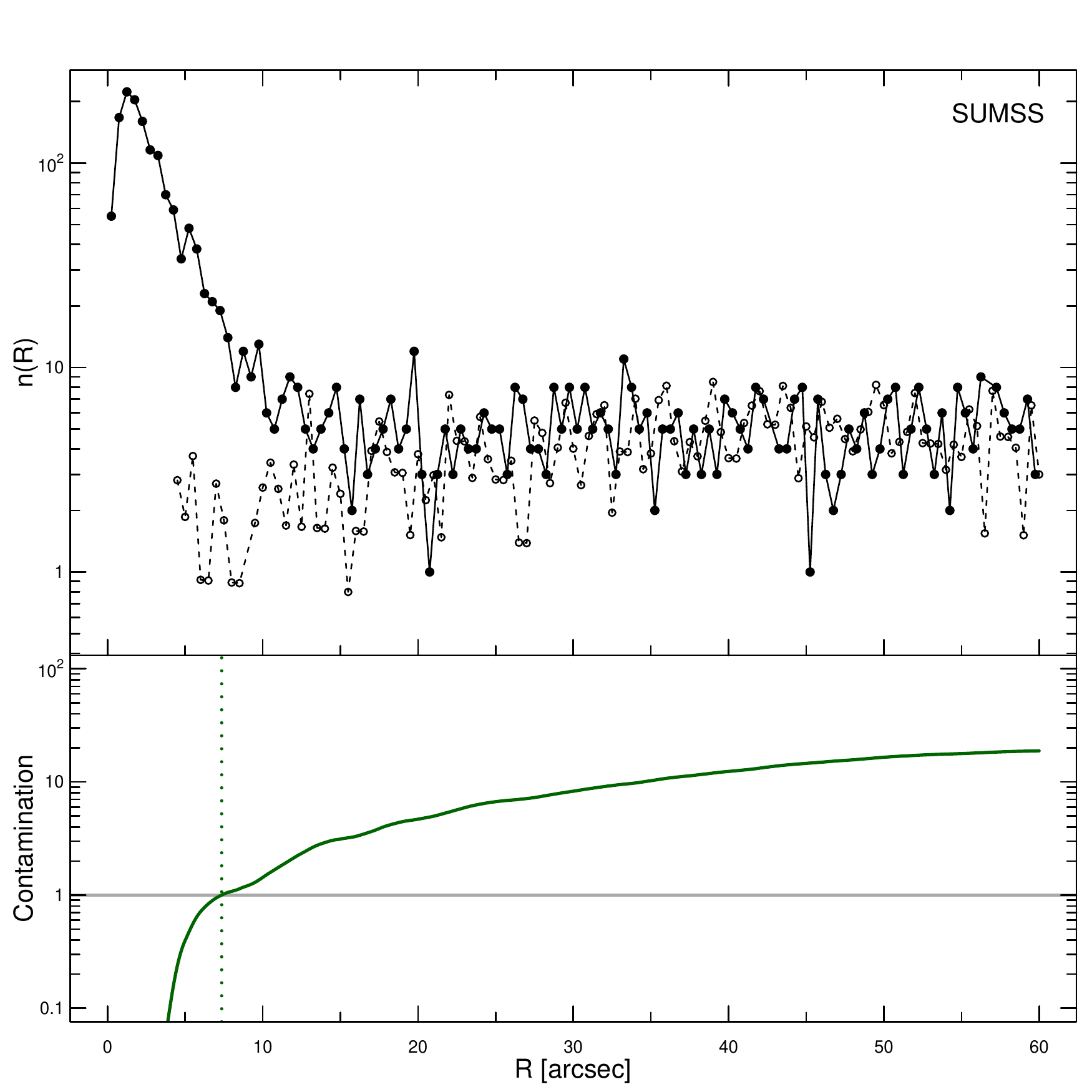}	
	\includegraphics[height=7cm,width=6cm,angle=0]{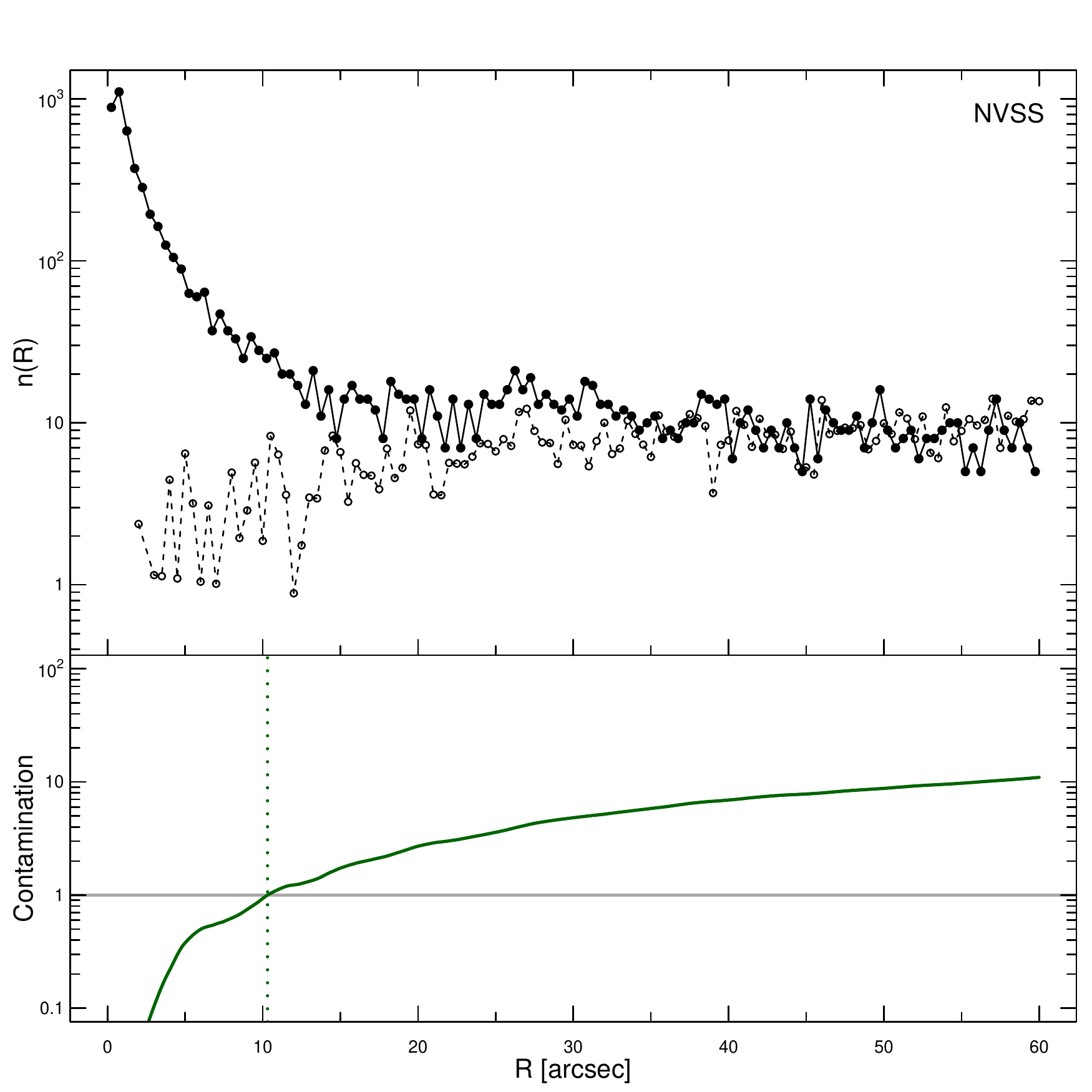}	
	\includegraphics[height=7cm,width=6cm,angle=0]{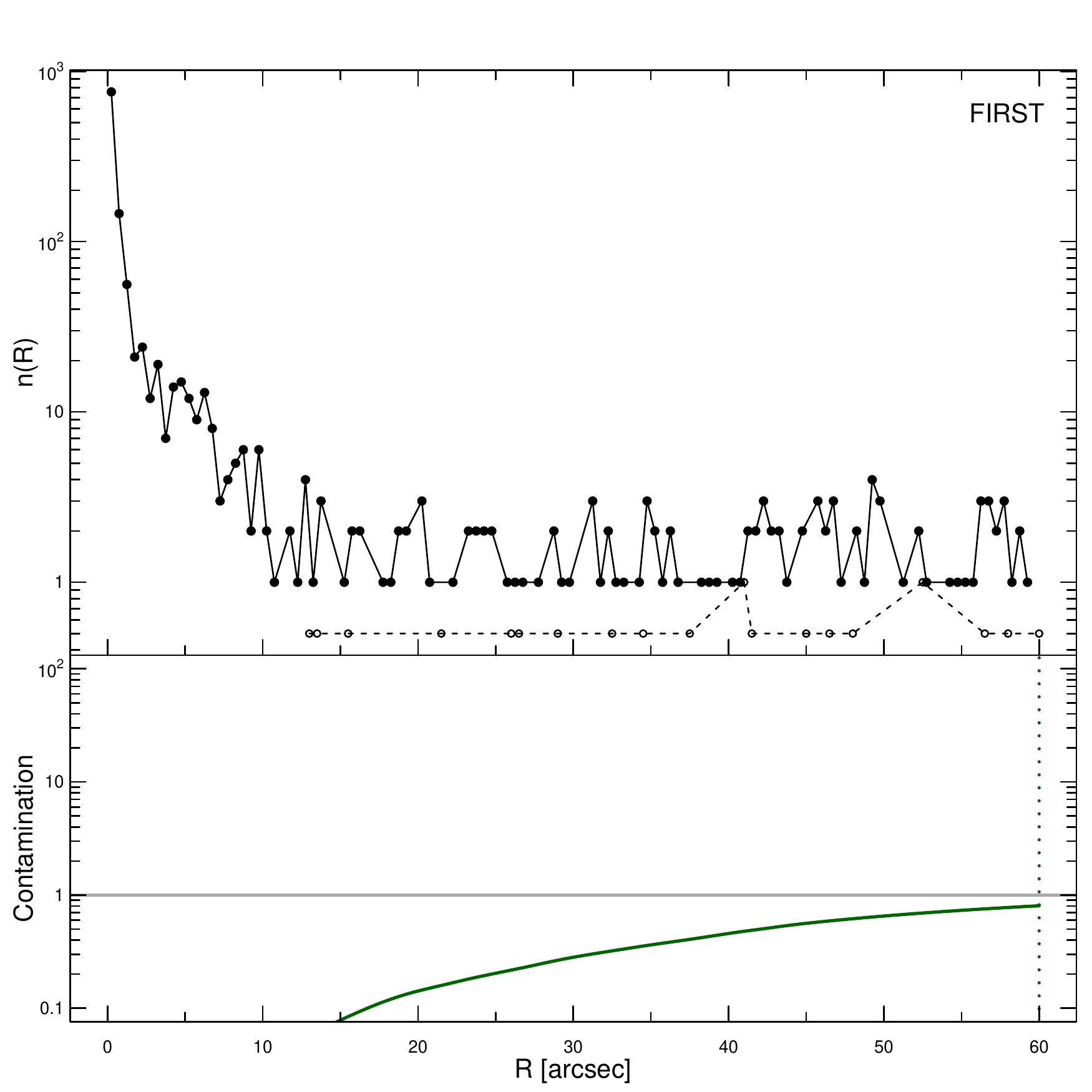}
    	\caption{Upper panels: number of real cross-matches (solid lines) and mock 
	cross-matches (dotted lines) for WISE sources detected in all bands
	as function of the radial distance $r$ in the interval $[0\arcsec, 60\arcsec]$, 
	for SUMSS, NVSS and FIRST surveys (from left to right). Lower panels: contamination of the 
	selection procedure for radial distances in the interval $[0\arcsec, 60\arcsec]$ for each 
	of the three radio surveys used in this paper. The horizontal lines show the contamination 
	threshold $c_{\mathrm{thr}}\!=\!1\%$ and the 
	vertical lines indicate the optimal radii for NVSS and SUMSS surveys, chosen as described 
	in Section~\ref{subsec:spatial_selection}. For FIRST survey, the procedure used to select the optimal
	crossmatch radius employed in this paper is different (see Section~\ref{subsec:spatial_selection}).}
         \label{fig:radii}
\end{figure*}

We have estimated $n^{\mathrm{real}}(r)$ by counting the number of WISE sources detected in all four
bands within circular regions of radius $r$ between 0\arcsec and 60\arcsec\, centered on a sample 
of $5\!\cdot\!10^{4}$ sources randomly extracted from each of the three radio surveys. In order to estimate 
the corresponding $n^{\mathrm{mock}}(r)$ values, 
we have created one hundred mock 
realizations of the coordinates of each real radio source by shifting its position by a radial distance randomly 
drawn from the interval $[60, 120]$\arcsec\ in a random direction. The numbers of WISE sources 
associated to real and mock radio positions are shown in the upper panels of Figure~\ref{fig:radii} as 
functions of the radial distance (real and mock
crossmatches are shown as solid and dashed black curves respectively). The numbers of crossmatches 
around mock positions can be fractional as they have been averaged over the 100 mock realizations of the 
real radio positions. The lower panels in Figure~\ref{fig:radii} show the contamination $c(r)$, calculated 
with the equation~\ref{eq:contamination}, as a function of the radial 
distance from the radio coordinates. In the lower panels the horizontal line corresponding to the threshold
contamination $c_{\mathrm{thr}}\!=\!1\%$ (efficiency $e(r)\!=\!99\%$) and the vertical lines indicate the 
optimal radii, corresponding to the intersection of the horizontal lines with the completeness curve.

The optimal cross-match radii determined with this method are $r_{\mathrm{NVSS}}\!=\!10.3\arcsec$, 
$r_{\mathrm{SUMSS}}\!=\!7.4\arcsec$ for the NVSS and SUMSS surveys, respectively. The number of 
WIBRaLS sources with distinct NVSS and SUMSS radio counterparts selected, as a consequence, is
11928 and 2325, respectively. 

Our method does not produce a reasonable estimate of the optimal radius for the FIRST survey 
(see right plot in Figure~\ref{fig:radii}), likely because of the very high density of FIRST sources
in the sky compared to the other two surveys.
The black solid line in the right plot of Figure~\ref{fig:radii} declines very steeply at small radial distances, 
with $\sim\!85\%$ of the total crossmatches found at distances smaller than $5\arcsec$. Based on this
fact, we have adopted a different approach to determine a crossmatch radius for FIRST.
We have chosen as optimal radial distance three times the combined positional uncertainties of 
AllWISE and FIRST detections. The maximum allowed positional uncertainty of AllWISE 
detections along each axis is 0.5\arcsec~\footnote{See\\
http://wise2.ipac.caltech.edu/docs/release/allwise/expsup/sec2$\_$5.html\\ for details}, even 
though for most of the sources detected the error is $\sim\!0.02\arcsec$ per axis, yielding 
a total positional uncertainty of $\sim\!0.1\arcsec$. In order to be conservative in our analysis, we have 
assumed a positional uncertainty for AllWISE sources $\sigma_{\mathrm{W1}}\!=\!0.5\arcsec$.
The astrometric accuracy of FIRST radio sources down to the survey flux threshold is consistently
better than 1\arcsec~\citep{white1997}. Nonetheless, also in this case we will assume a conservative 
positional uncertainty $\sigma_{\mathrm{FIRST}}\!=\!1\arcsec$ for FIRST detections.
Thus, the combined positional uncertainty can be estimated as
$\sigma_{\mathrm{W1}\!+\!\mathrm{FIRST}}\!=\!\sqrt{\sigma_{\mathrm{W1}}^2\!+\!\sigma_{\mathrm{FIRST}}^2}\!\sim\!1.12\arcsec$, which yields an optimal radius for FIRST sources
$r_{\mathrm{FIRST}}\!=\!3\sigma_{\mathrm{W1}\!+\!\mathrm{FIRST}}\!=\!3.4\arcsec$. 
Using this maximum radial distance, we select 106 WIBRaLS sources with FIRST radio counterparts  
that cannot be associated to a NVSS source.

We do not attempt to estimate the completeness of our selection procedure because 
the properties of the parent population of our sample of WIBRaLS are impossible to determine. In the 
case of the FIRST survey, the adoption of a smaller maximum radial distance compared to the
radii used for NVSS and SUMSS surveys, possibly results in a lower completeness. 
However, we estimate the loss of real crossmatches to be small: the increase in the number of 
sources selected with our procedure 
using radial distance $r\!=\!10\arcsec$ is only 104, corresponding to $\sim\!10\%$ of the number of 
WIBRaLS selected with the optimal radius $r_{\mathrm{FIRST}}\!=\!3.4\arcsec$.

The number of WISE blazar-like sources with at least one radio counterpart within the maximum 
radial distances discussed above is 11928, 6592 and 2325 for NVSS, FIRST and SUMSS
surveys, respectively (see Table~\ref{tab:samples}), for a total of 20845 WISE blazar-like 
sources spatially associated to distinct radio counterparts from either of these three catalogs.
We find that 1552 WISE blazar-like sources are associated to one NVSS and one SUMSS radio 
source, while 3660 WISE blazar-like sources are cross-matched to one NVSS and one FIRST 
radio sources. For this reason, the final number of unique WISE blazar-like sources with at least 
one radio counterpart is 16632.

\subsection{Radio-loudness Selection}
\label{subsec:q22_selection}

Blazars are by definition radio-loud AGNs. For this reason, in our analysis 
we only consider WISE blazar-like sources that can be spatially associated to at least one radio 
counterpart in one of the three radio surveys within the maximum radial distance discussed 
in Section~\ref{subsec:spatial_selection}. However, radio emission can be also produced 
by physical mechanisms not due to the presence of an AGN. For example, it is well known 
that the far-IR and radio emission are tightly and linearly correlated in star-forming 
systems~\citep[see e.g.,][and references therein for more details]{sargent2010,bonzini2012}.
The strength of the correlation between radio and far-IR emissions is usually expressed via the 
so-called $q$ parameter, defined as the logarithm of the ratio of far-IR to
radio flux density, \citep[e.g.,][]{helou1985}. Unfortunately, flux density measurements at 
far-IR frequencies required to compute the $q$ parameter are often not available. 
However,~\cite{padovani2011} and~\cite{bonzini2013} have recently shown that it is possible to 
define a $q_{24}$ parameter as:

\begin{equation}
	q_{24}\!=\!\log{(S_{24\mu m}/S_{1.4\mathrm{GHz}})}
\end{equation} 

\noindent where $S_{24\mu m}$ is the observed flux density at 24$\mu$m and $S_{1.4GHz}$ is the 
flux density measured at 1.4 GHz. The use of the observed flux densities minimizes the uncertainties 
introduced by the modeling of the spectral energy distribution~\citep{bonzini2013}. It is worth noting 
that the 24$\mu$m band of the Multi-band Imaging Photometer for {\it Spitzer} 
(MIPS), used to measure the $S_{24\mu m}$ in the previous analyses, is similar to the WISE $[22] \mu$m
band\footnote{http://wise2.ipac.caltech.edu/docs/release/allsky/expsup/sec6$\_$3a.html}~\citep{wright2010,cutri2011}.
The passbands of the WISE $[22]$ and MIPS-24 band are similar, although the $[22]$ WISE filter is 
slightly bluer in response\footnote{http://wise2.ipac.caltech.edu/docs/release/prelim/expsup/sec4$\_$3g.html}.
For all the WISE blazar-like sources associated to one radio source using the method described in 
Section~\ref{subsec:spatial_selection}, we calculated the $q_{22}$ parameter as follows:
 
\begin{equation}
	q_{22}\!=\!\log{(S_{22\mu m}/S_{\mathrm{radio}})}
\end{equation} 

\begin{figure}[]
 	\includegraphics[height=6cm,width=8.5cm,angle=0]{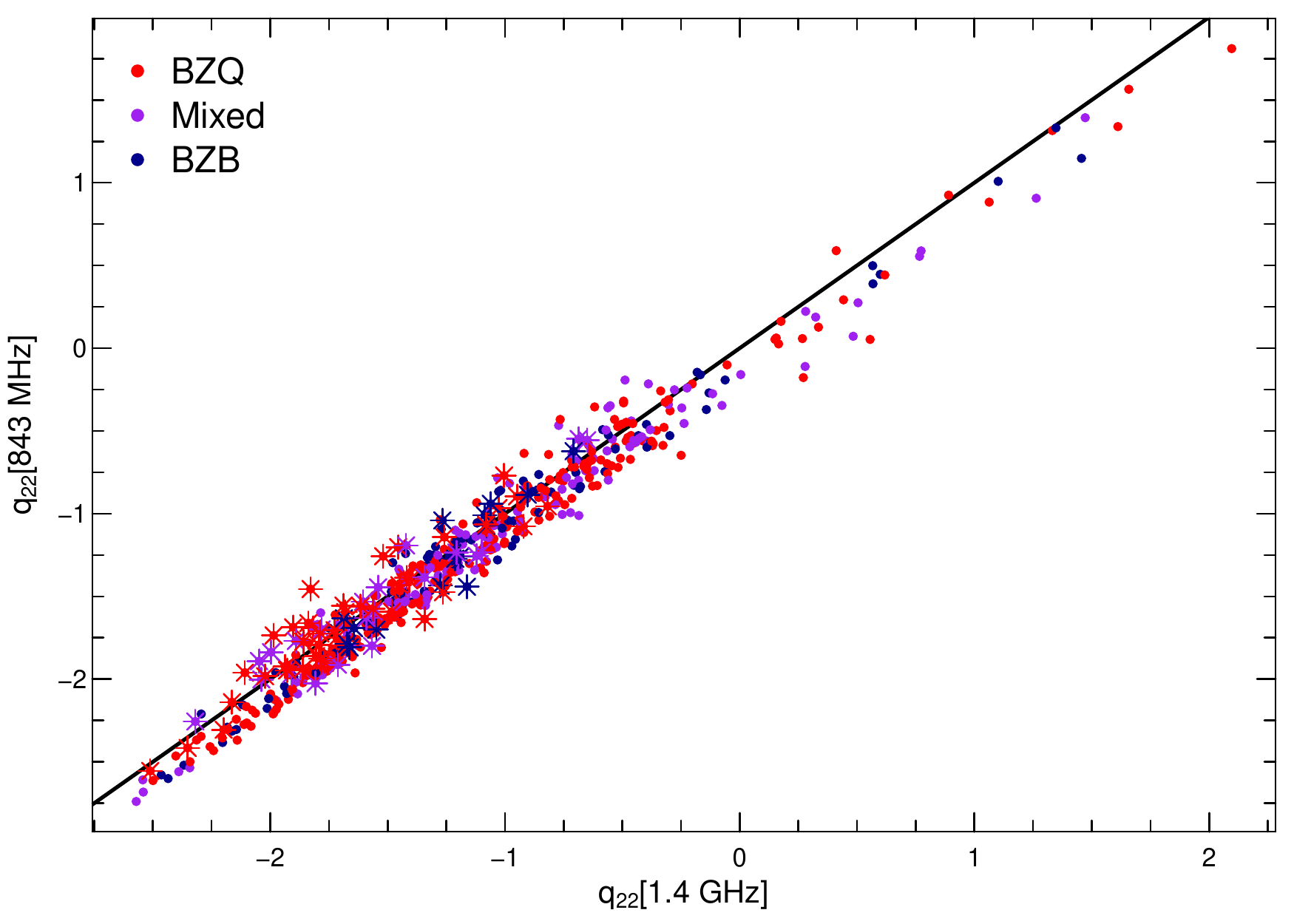}
	\caption{Values of the $q_{22}[1.4~\mathrm{GHz}]$ and $q_{22}[843~\mathrm{MHz}]$ parameters
	for all WISE blazar-like sources associated to a radio counterpart in both the NVSS and SUMSS 
	surveys. The stars represent the 73 confirmed blazars in this sample that can be cross-matched
	to a BZCat counterpart. The spectral classification of the WISE blazar-like sources based on their WISE colors 
	(see Section~\ref{subsec:wise_selection}) is color-coded.}
        \label{fig:q22_nvss_sumss}
\end{figure}

Following~\cite{bonzini2013}, we used the flux density at 1.4 GHz as the radio flux density 
$S_{\mathrm{radio}}$. Since flux density measurements at 1.4 GHz are not available 
in the SUMSS survey, for the SUMSS counterparts we used the flux 
densities at 843 MHz instead. A well known property of the blazars is the flatness of 
their radio spectra~\citep[see e.g.,][]{healey2007} that extends up to low radio frequencies 
well below 1 GHz~\citep[see also][for a recent discussion]{massaro2013a,massaro2013c,massaro2013d,massaro2014}.
For this reason, the use of the flux density at 843 MHz instead of the flux density at 1.4 GHz 
to estimate the $q_{22}$ parameter affects negligibly our analysis.
Nonetheless, we checked that the differences introduced in the value of the parameter $q_{22}$ by 
the use of the flux density measured at 843 MHz instead of the flux density at 1.4 GHz are small. 
We used the 553 WISE blazar-like sources with one radio counterpart in the SUMSS catalog 
and another in the NVSS catalog (see Section~\ref{subsec:spatial_selection}). For these
sources we calculated $q_{22}$ parameter values using both
flux densities measured at 1.4 GHz and 843 MHz. Figure~\ref{fig:q22_nvss_sumss} shows the 
$q_{22}(1.4~\mathrm{GHz})$ vs $q_{22}(843~\mathrm{MHz})$ distribution for this sample of sources. 
The difference between the values of $q_{22}(1.4~\mathrm{GHz})$ and $q_{22}(843~\mathrm{MHz})$ 
for all the WISE blazar-like sources with radio counterparts in the two surveys is smaller than 10\% of 
the $q_{22}(1.4~\mathrm{GHz})$ value for $\sim88\%$ of the sources. 
We also notice that the distribution of the differences between the values of the two values of $q_{22}$ 
peaks at $\Delta q_{22}\!=\!q_{22}(843~\mathrm{MHz})\!-\!q_{22}(1.4~\mathrm{GHz})\!\sim\!0.08$ and 
does not depend on the WISE spectral class of the sources considered.
Moreover, $\Delta q_{22}$ is almost constant over the range of $q_{22}$ spanned by our sample of WISE 
blazar-like sources with radio counterparts in both the NVSS and SUMSS surveys. For this reason, for 
radio-loud WISE blazar-like radio sources with only SUMSS radio counterpart, we have used the corrected 
$q^{'}_{22}(843~\mathrm{MHz})\!=\!q_{22}(843~\mathrm{MHz})\!+\!\Delta q_{22}$ where 
$\Delta q_{22}\!=\!0.08$. The change in the number of 
sources selected using the corrected $q^{'}_{22}$ is anyway $\sim\!3.5\%$ of the total sample of WIBRaLS 
(see Section~\ref{sec:catalog}). It is worthwhile to stress that the scatter of the two $q_{22}$ estimates for the 
subset of 73 candidate blazars in this 
sample associated to a confirmed blazar of the ROMA-BZCat catalog within 3.3\arcsec 
(stars in Figure~\ref{fig:q22_nvss_sumss})
is smaller ($\sigma_{\Delta q_{22}}\!=\!0.12$) and their distribution is less biased 
($<\Delta q_{22}> =\!0.01$) than the distribution 
of the whole sample. Nonetheless, we have used the correction 
$\delta q_{22}\!=\!0.08$ derived from the whole sample because the nature of most WISE blazar-like 
sources with radio counterparts has not been confirmed yet. 

\begin{figure}[] 
	\includegraphics[height=8cm,width=8.5cm]{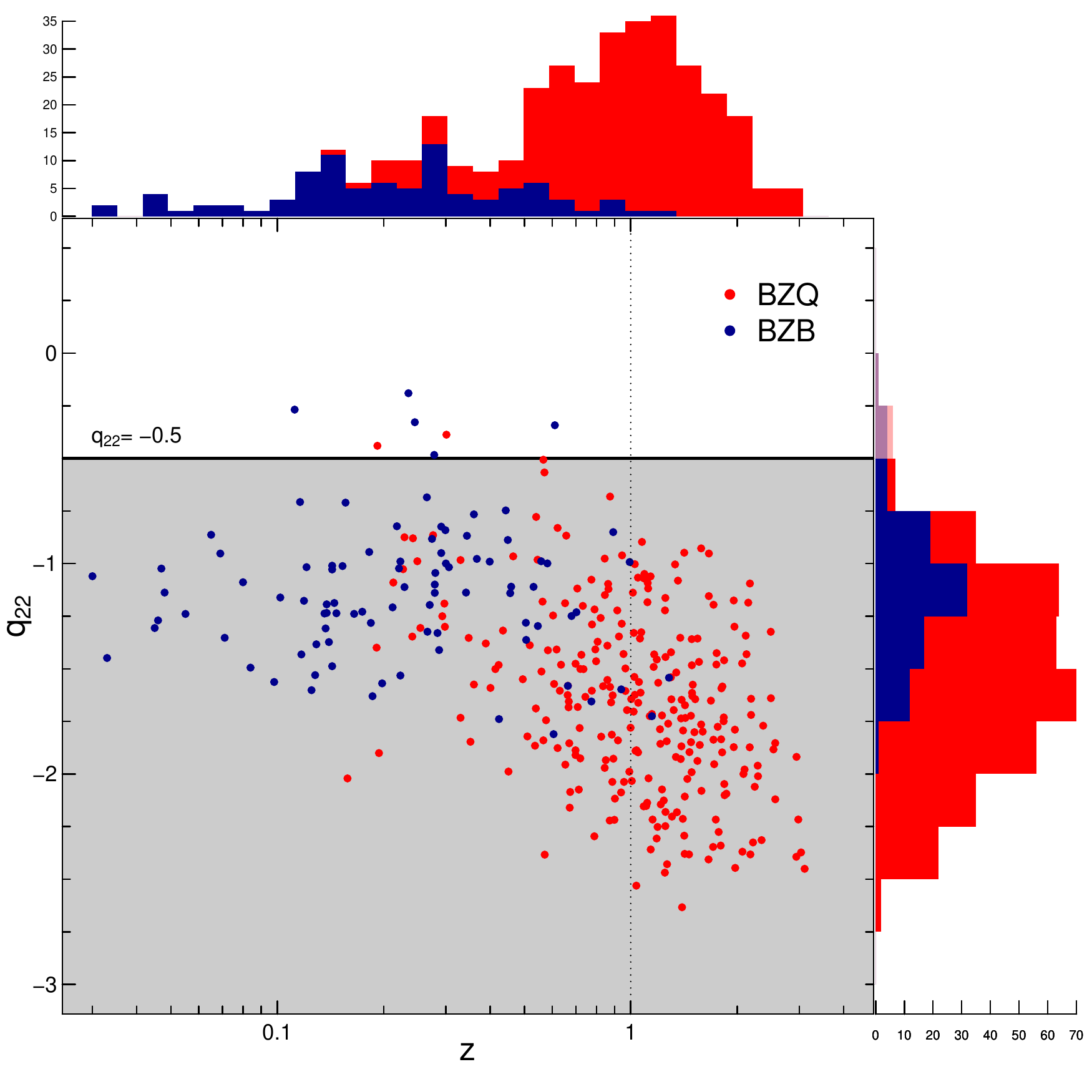}
	\caption{Scatterplot of the $q_{22}$ values for the confirmed $\gamma$-ray emitting
	blazars in the {\it locus} sample as a function of their redshifts (from the BZCat), with
	marginal histograms ({\it locus} sources with undetermined or uncertain redshifts are excluded). 
	The black solid line shows the threshold $q_{22}\!\leq\!-0.5$ used to select the WIBRaLS 
	(see discussion in Section~\ref{subsec:q22_selection}.}
         \label{fig:q22_locus}
\end{figure}

\cite{bonzini2013} showed that the $q_{24}$ parameter has a redshift dependence (see~Figure 2 
in~\citealt{bonzini2013}): the $q_{24}$ values of all classes of sources considered (radio-loud AGNs, 
radio-quiet AGNs and star-forming galaxies) become smaller for larger redshifts. For this reason, the 
boundary between the regions of the redshift vs $q_{24}$ plane (Figure~2 in~\citealt{bonzini2013})
dominated by the radio-loud AGNs and the other radio sources is also function of the redshift. In principle, 
if redshift estimates for all WISE blazar-like sources associated to a radio counterpart were
available, we could have removed the redshift dependence and selected radio-loud WISE blazar-like sources 
with a radio counterpart at each redshift. Since redshifts are not available, we can only set a fixed threshold 
for the $q_{22}$\footnote{The determination of the redshifts of BZBs can be
intrinsically difficult, because of the lack of significant features in their optical spectra. In order to increase
the number of confirmed blazars with reliable spectroscopy, we are carrying out
an extensive observational program to acquire the spectra of a large number of WISE-selected 
candidate blazars. The first results are discussed in~\cite{paggi2014}}.
Figure~\ref{fig:q22_locus} shows the distribution of $q_{22}$ values calculated for the confirmed 
$\gamma$-ray emitting blazars in the {\it locus} sample as a function of the redshifts and color-coded 
according to their spectral classification from the BZCat. We have 
excluded the sources of the {\it locus} (mostly BZBs) whose redshifts are not well determined or unknown.  
All {\it locus} sources have $q_{22}$ lower than 0 and $\sim\!96.5\%$ have $q_{22}\!\leq\!-0.5$ 
($\sim\!94\%$ for sources classified as BZB - blue points - and $\sim\!99\%$ for sources classified as 
BZQ - red points). Based on these observational evidences, we will require WIBRaLS sources to 
have $q_{22}\!\leq\!-0.5$ in order to minimize the contamination from radio-quiet AGNs which can have 
similar WISE colors, at the cost of decreasing the overall completeness of the selection by less than $5\%$.
\footnote{We have also evaluated the effect of the redshift distribution on the distribution of the $q_{22}$ values 
of the WISE blazar-like sources
and on the fraction of sources
selected by the $q_{22}\!\leq\!-0.5$ condition. We have calculated the $q_{22}$ values for each 
confirmed $\gamma$-ray emitting blazars in the {\it locus} sample 
after varying its redshift on a regularly spaced grid in the interval $[0, 4]$, 
and assuming a power-law spectral energy distribution with slope defined by the observed flux densities
at $22\mu$m and at 1.4GHz. We found that $\sim\!94\%$ of the estimated $q_{22}$ values for all redshift
values are still smaller than -0.5.}
 
It is interesting to discuss how $q_{22}\!\leq\!-0.5$ condition affects the completeness
of our selection. By assuming completeness limits of $\sim\!2.5$ mJy at 1.4 GHz for NVSS 
catalog~\citep{condon1998}, $\sim\!1$ mJy at 1.4 GHz for FIRST~\citep{becker1995} 
and $\sim\!8$ mJy at 843 MHz~\citep{mauch2003}, and with a flux limit $\sim\!6$ mJy in 
the WISE $[22]$ filter (coverage depth 11, SNR$\!=\!11$)~\footnote{http://wise2.ipac.caltech.edu/docs/release/allwise/expsup/sec2$\_$3a.html}, the nominal value for our selection $q_{22}\!\leq\!-0.5$ (see 
Figure~\ref{fig:q22_locus}) at the WISE $22\mu$m limit therefore implies a flux density in the radio 
$\geq\!20$ mJy, within the detection limit of all the radio catalogs. Therefore our WISE-selected 
sample should be complete in this regard. On the other hand, some fainter blazars detected 
in the radio may have escaped detection with WISE in the $[22]$ filter, and the BZB class 
would be most likely to suffer from this deficit because, typically, the synchrotron peak of their spectral energy
distributions is at higher energies than FSRQs. Future deep mid-IR followups, perhaps with the 
James Webb Space Telescope, might extend the sample of {\it bona fide} blazars with measured mid-IR 
properties.
 
We have explored the possibility that Steep-Spectrum Radio Quasars (SSRQs) contaminate 
the sample of WIBRaLS selected with our method. The SSRQs are 
powerful radio sources characterized by radio spectral index $\alpha_{R}\!>\!0.5$, usually calculated
between 1.4 GHz and 4.85 GHz. In order to evaluate the contamination 
from SSRQs, we have used a sample of 18 {\it bona fide} SSRQs selected by~\cite{gu2013} among the 
SDSS optical 
quasars in the Stripe 82 region and with radio counterparts in the FIRST, PMN and GB6 surveys. 
The crossmatch of the SSRQs radio positions with the WISE
AllWISE catalog within a maximum radius 3.4\arcsec (the cross-match radius for FIRST radio counterparts
estimated in Section~\ref{subsec:spatial_selection}) yields 18 unique WISE counterparts. 
Two of these WISE counterparts not detected in the W4 band have been discarded. The application of 
the WISE colors selection method described in Section~\ref{subsec:wise_selection} to the 16 remaining
WISE SSRQs counterparts produced 11 sources selected as WISE blazar-like sources, all classified as
BZB-type candidates. The distribution 
of the $q_{22}$ values for the WISE counterparts of the 11 SSRQs spans the interval $[-2.2, 0.3]$, and 8 
of them ($\sim\!73\%$) have $q_{22}\!\leq\!-0.5$, i.e. are compatible with the $q_{22}\!\leq\!-0.5$ condition 
used to extract the 
WIBRaLS. Therefore, 50\% of the SSRQs sample produced by~\cite{gu2013} is included in the WIBRaLS catalog. 
We found that the WIBRaLS catalog can contain SSRQs and that it is not possible to 
selectively exclude this class of sources using 
the $q_{22}$, since their $q_{22}$ values have similar distribution to the $q_{22}$ 
values for the confirmed $\gamma$-ray emitting blazars. While a quantitative estimate of the contamination 
from SSRQs contaminants based on this small sample is impossible, we discuss this point further in 
Section~\ref{subsec:veroncat}.

\begin{figure}[] 
	\includegraphics[height=8cm,width=8.5cm]{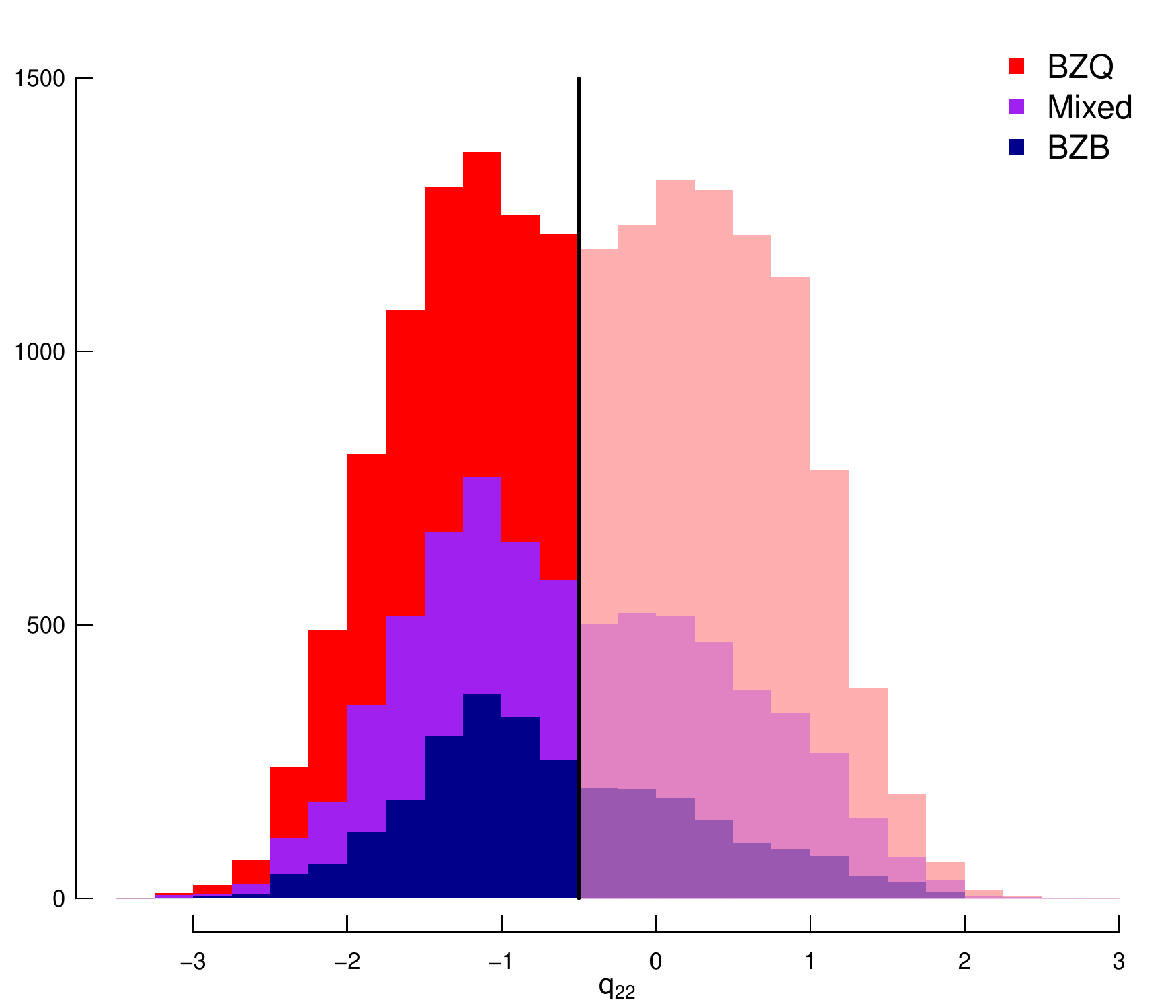}	
	\caption{Histogram of the distribution of the $q_{22}$ values for all the WISE blazar-like
	sources associated to a radio counterpart in one of the surveys used in this paper (BZB-type candidates
	are blue, BZQ candidates red and sources associated to the Mixed region are magenta). The full 
	colors show the histogram of the final catalog of WIBRaLS, obtained with 
	$q_{22}\!\leq\!-0.5$.}
         \label{fig:q22_histogram}
\end{figure}

The distribution of the $q_{22}$ values for all sources in the catalog of WISE blazar-like sources
with at least one radio counterpart, color-coded according to the WISE spectral classification in 
BZB-type, BZQ-type candidates and sources compatible with the Mixed region of the {\it locus} model, 
is shown in Figure~\ref{fig:q22_histogram}. The solid colors represent the sources selected as 
WIBRaLS based on the criterion $q_{22}\!<\!-0.5$. The distribution of $q_{22}$ values for the 
sources classified as BZB-type candidates and Mixed show only one peak located in the 
$-1.5\!<\!q_{22}\!<\!-0.5$ range, while the distribution of $q_{22}$ values for sources classified as 
BZQ is clearly bimodal, with another peak at $q_{22}\!\sim\!0.5$.

The number of total distinct WISE blazar-like sources with a radio counterpart selected as WIBRaLS
is 10002. Using the $q_{22}\!\leq\!-0.5$ criterion, we selected 6362, 1775 and 1865 distinct 
radio sources in the NVSS, FIRST and SUMSS surveys respectively (see Table~\ref{tab:samples}). 
The final number of unique WIBRaLS is 7855 (see Table~\ref{tab:wibrals_unique}). 
In Section~\ref{sec:catalog}, we provide more details on the composition of the 
final catalog of WIBRaLS.

\section{The WIBRaLS Catalog}
\label{sec:catalog}

We applied the three-steps procedure described in Section~\ref{sec:selection} to the whole WISE 
AllWISE Catalog of sources detected in all four WISE filters, selecting a total of 
$\sim\!2.65\cdot10^{5}$ WISE blazar-like sources (the composition of this sample in terms of 
WISE classes and types is shown in Table~\ref{tab:samples}). Then, we selected the WISE blazar-like 
sources that can be spatially associated to one radio source from three radio surveys, namely the NVSS, 
SUMSS and 
FIRST surveys, using the optimal radii established in Section~\ref{subsec:spatial_selection}. 
We found a total of 20845 WISE blazar-like sources with at least
one counterpart from either one of the three radio surveys. In particular, 11928, 6592 and 2325 
WISE blazar-like sources have a unique radio counterpart in the NVSS, FIRST and 
SUMSS surveys, respectively. 3660 WISE blazar-like sources can be associated to one source in 
both the NVSS and FIRST surveys and 553 WISE blazar-like sources can be associated to one radio
source in both the NVSS and SUMSS surveys, respectively. For consistency, in the case of multiple
radio counterparts from distinct surveys, we have always adopted as final radio counterpart of 
the WISE blazar-like source the NVSS source (FIRST and SUMSS footprints do not overlap), because
NVSS covers the largest area among the radio surveys used in this paper.
The final list of single distinct radio counterparts of our sample of WISE blazar-like sources 
contains 11928 NVSS sources, 2932 FIRST sources and 1772 SUMSS sources (see 
Table~\ref{tab:samples}), for a total of 16632 WISE blazar-like sources with a unique radio 
counterpart. We finally extract the catalog of WIBRaLS by selecting only the
WISE blazar-like sources with a radio counterpart with radio-loudness parameter $q_{22}\!\leq\!-0.5$
(see Section~\ref{subsec:q22_selection} for details on the $q_{22}$ parameter and the selection performed).

\begin{figure*}[]
 	 \begin{centering}
	 \includegraphics[scale=0.52]{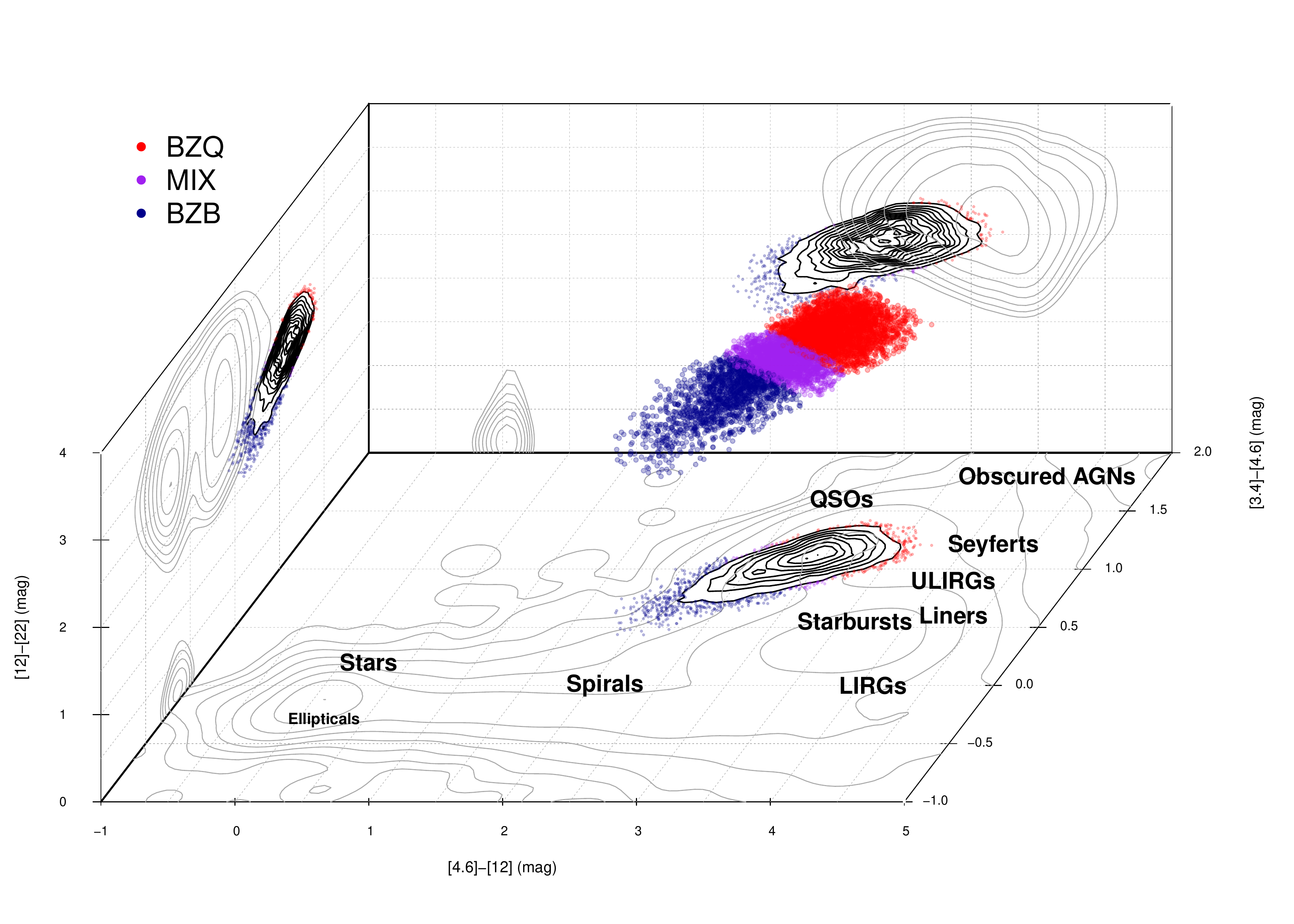}
	 \caption{Three dimensional distribution of the sources in the WIBRaLS catalog (each
	 source is color-coded according
	 to the WISE spectral classification) in the space generated by the WISE $[3.4]\!-\![4.6]$, $[4.6]\!-\![12]$ 
	 and $[12]\!-\![22]$ colors. The black and gray lines displayed on the three planes represent 
	 the projected isodensity contours associated to ten log-spaced levels of the 
	 three-dimensional distribution of WIBRaLS and of a sample of random sources detected in 
	 all four WISE filters, respectively. The approximate locations 
	 of different typical classes of objects in the $[4.6]\!-\![12]$ vs $[3.4]\!-\![4.6]$ color-color plane, 
	 according to~\cite{wright2010}, are also shown.}
	 \label{fig:3dcolors}
	 \end{centering}
\end{figure*}

The final catalog of WIBRaLS contains 7855 unique sources that satisfy all our criteria: 
6362 of these WISE sources are associated to a NVSS counterpart, 1407 to a source 
in the SUMSS survey and the remaining 86 can be cross-matched to a unique 
source from the FIRST survey (Table~\ref{tab:wibrals_unique}). The final WIBRaLS sample, 
according to the spectral classification based on the WISE colors and discussed in 
Section~\ref{subsec:wise_selection}, is composed 
of 1682 sources classified as BZB, 3973 sources 
classified as BZQ and 2194 sources whose colors are compatible with
the Mixed region of the {\it locus} model. The WIBRaLS can also be split in 
129 class A sources ($\sim\!2.4\%$ of the total sample, 714 class B sources 
($\sim\!9.1\%$) and 7012 class C sources ($\sim\!89\%$). These fractions are
similar to the class fractions for the sample of WISE blazar-like sources before the 
selection based on the radio counterparts and $q_{22}$ radio-loudness parameters 
(Section~\ref{subsec:wise_selection}). They follow from the stringent definition 
of Class A (see Section~\ref{subsec:wise_selection}) in terms of the threshold on the 
WISE {\it score}. The composition of the final WIBRaLS catalog is shown in 
Table~\ref{tab:wibrals_unique}. 

Among the 7855 WIBRaLS, 1295 sources ($\sim\!16.5\%$) 
can be spatially re-associated to a known blazar listed in the ROMA-BZCAT 
(v4.1)\footnote{http://www.asdc.asi.it/bzcat/}~\citep{massaro2009,massaro2011} 
within 3\arcsec.3 \citep[see also][for the choice of this association radius]{dabrusco2012,dabrusco2013}, 
corresponding to $\sim\!41\%$ of the 3149 total members of the ROMA-BZCat. Moreover, 
454 of the 1295 ROMA-BZCat counterparts of the WIBRaLS can be spatially associated 
to a $\gamma$-ray emitting source of the {\it locus} sample ($\sim\!76.5\%$), extracted from 
the 2FGL catalog~\citep{nolan2012} of $\gamma$-ray sources. This fraction is lower than 
the 81\% fraction of {\it locus} sources contained, by definition, in 
the model of the {\it locus} in the WISE colors space~\citep{dabrusco2013}. The 
fraction of confirmed ROMA-BZCat sources in the {\it locus} sample that can be spatially associated 
to one member of the WIBRaLS catalog ($\sim\!76.5\%$) is also  
significantly lower than the 96.5\% of {\it locus} sources selected by directly applying 
the condition on the $q_{22}$ value to their radio and WISE counterparts 
(see Section~\ref{subsec:q22_selection}).
The reason for the discrepancy between the 81\% of the {\it locus} sample contained
in the {\it locus} model by definition, and the $\sim\!76.5\%$ of {\it locus} sources 
that are selected as WIBRaLS can be found in the significant differences between 
the WISE photometry of $\sim\!10\%$ of the {\it WISE counterparts of the locus} sources in 
the AllSky release used to define the {\it locus} 
model by~\cite{dabrusco2013} and the photometry in the AllWISE WISE release, 
used here.

The distribution of the final catalog of WIBRaLS in the three dimensional space generated by the 
WISE colors $[3.4]\!-\![4.6]$, $[4.6]\!-\![12]$ and $[12]\!-\![22]$, is shown in Figure~\ref{fig:3dcolors}, 
where each source is color-coded according to the spectral class assigned based on its WISE colors. 
The three projections of the WIBRaLS 3D WISE colors distribution onto the three color-color planes 
clearly show that, in each plane, the peaks of the density distributions lie in the regions 
occupied by the sources classified as BZQ. Moreover, in Figure~\ref{fig:3dcolors}, $10^5$ 
generic WISE sources located at high galactic latitude ($|b|\geq\!15^{\circ}$) and detected in all 
four filters have been used to plot the gray arbitrary log-spaced isodensity contours. The 
comparison with the black density contours
of the WIBRaLS dataset shows that significant overlap between the distributions of WIBRaLS and 
generic WISE sources in each of the three color-color diagrams can only removed in the 3D WISE 
colors space.

\begin{table*}
	\small
	\begin{center}
	\caption{Total number of sources in the catalog of WISE blazar-like sources (left table), the catalog of WISE blazar-like 
	with radio counterparts (mid), and the catalog of WIBRaLS ($q_{22}\!\leq\!-0.5$,
	right table), split for WISE class and type (see Section~\ref{subsec:wise_selection}). Where applicable, the numbers of 
	radio counterparts for each radio survey are also shown. The right side table shows the spectral classes and 
	WISE-based partition in classes for WIBRaLS sources (multiple radio counterparts of WISE blazar-like sources 
	are counted separately).}
	\begin{tabular}{lcccclccccc|cccc}
	\tableline
	\multicolumn{1}{c}{} & \multicolumn{4}{c}{WISE blazar-like sources} & &\multicolumn{4}{c}{WISE blazar-like sources} & & \multicolumn{4}{c}{WIBRaLS} \\
	\multicolumn{1}{c}{} & \multicolumn{4}{c}{} 		&  & \multicolumn{4}{c}{with radio counterparts} & &\multicolumn{4}{c}{} \\
	\tableline
			& BZB	& Mixed	& BZQ	&  Total 	& 		&BZB		& Mixed		& BZQ		& 	 Total 	&	& 	BZB		& Mixed		& BZQ		&  Total	\\
	\tableline
	Class A	&1345	&519		&1690	&3554		&		&  193		&236		&783		&	1212	&	&	45		& 30		&	86		& 161	\\
	Class B	&3985	&3611	&9904	&17500	&		&  536		& 635		&1663		&  	2834	&	&	244		&	220		&	476		&	 940	\\
	Class C	&27459	&58548	&158109	&244116	&		&  2694	&4583		&9522		&	16799	&	&	1813	&	2507	&	4581	&	 8901	\\
	 Total	&32789	&62678	&169703	&265170	& 		& 			& 			& 			&			&	& 			&			&			&			\\
	\cline{1-5}
			&		&		&		&			& NVSS	& 2093	& 3217	&6618		&	11928	&	& 1352	& 1804	&	3206	&	 6362	\\
			&		&		&		&			& FIRST	& 864		& 1615	& 4113	&	6592	&	& 349		&	468	&	958	& 1775	\\
			&		&		&		&			& SUMSS	& 466		& 622		& 1237	&	2325	&	& 401	& 485	&	979	&	 1865	\\
			&		&		&		&			&  Total&3423	&5454		&11968	&	20845	&	& 2012	& 2757	& 5143	&	 10002\\		
	\cline{6-15}
	\end{tabular}
	\end{center}
	\label{tab:samples}	
\end{table*}
 
\begin{table}
	\small
	\begin{center}
	\caption{Composition of the final WIBRaLS catalog (only the final radio counterparts 
	from the three radio surveys shown), in terms of WISE spectral
	types, classes and provenance of the radio counterpart (the NVSS sources have 
	always been considered as radio counterparts whenever available).}
	\begin{tabular}{lcccc}
	\tableline
			& 	BZB		& Mixed		& BZQ		& 	 Total	\\
	\tableline
	Class A	&37		& 26		&	66		& 129	\\
	Class B	&187		&165		&	362		& 714	\\
	Class C	&1458	&2003	&	3551		& 7012	\\
			&		&		&			&		\\
	NVSS	&1352	& 1804	&	 3206	& 6362	\\
	FIRST	& 21		&21		&	 44		& 86		\\
	SUMSS	& 309	& 369	& 	729		& 1407	\\		
	Total 	& 1682	&219		& 	3973		& 7855	\\
	\tableline
	\end{tabular}
	\end{center}
	\label{tab:wibrals_unique}	
\end{table}

\begin{figure*}[]
	 \includegraphics[height=12cm,width=16cm]{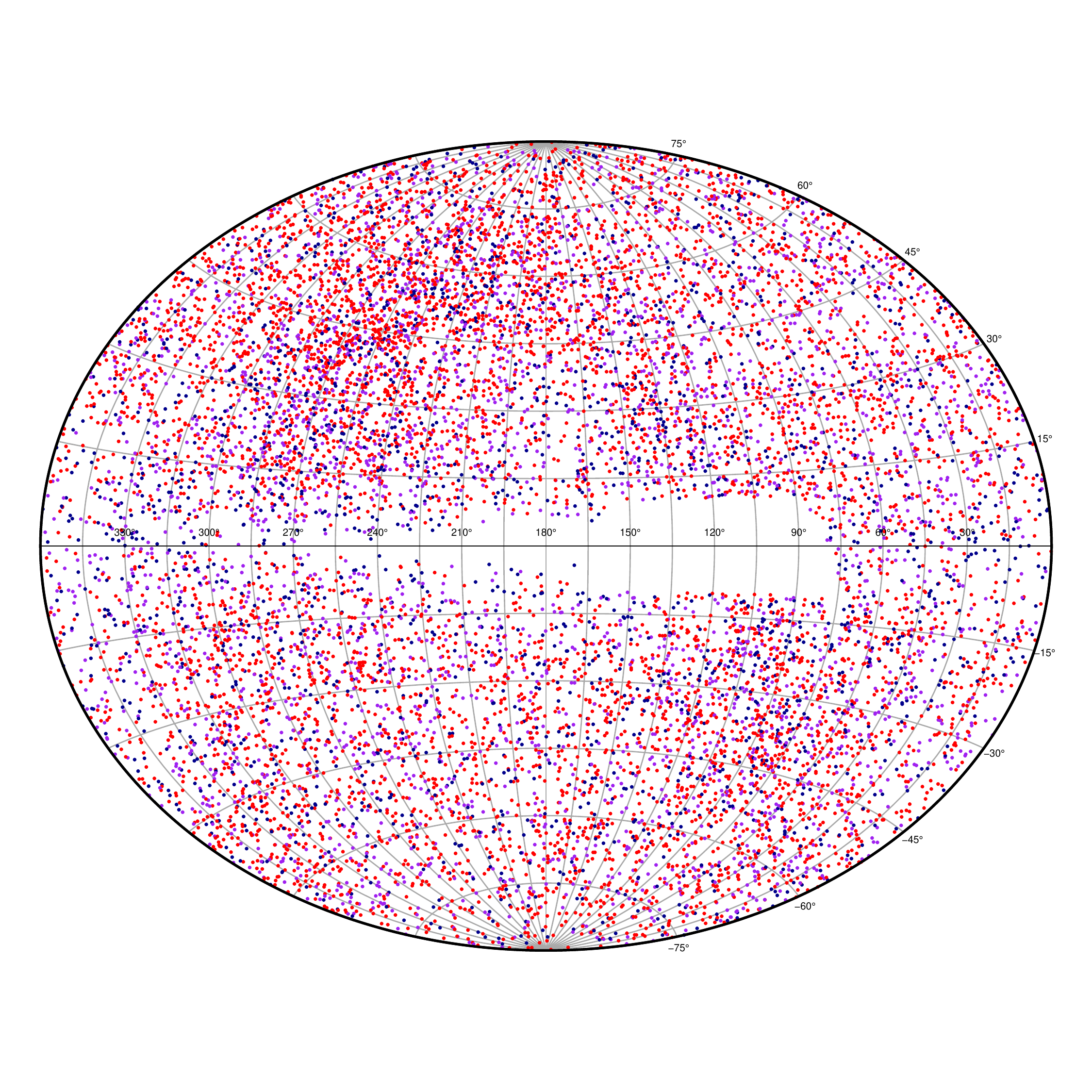}
	 \begin{centering}
	 \caption{Aitoff projection of the sky distribution in galactic coordinates of the catalog of
	 WIBRaLS, color-coded according to their WISE spectral classification in BZB, 
	 BZQ or Mixed. The empty region is not covered by any of the three radio surveys 
	 used in this paper to associate the WISE blazar-like sources.}
	 \label{fig:aitoff}
	 \end{centering}
\end{figure*}

The sky distribution in galactic coordinates of all WIBRaLS, 
color-coded according to their WISE spectral class, is shown in Figure~\ref{fig:aitoff}. The 
sky density of this sample depends on the spatial density of the 
sources in the WISE photometric catalog, whence the WISE sources with colors compatible with 
the {\it locus} of the $\gamma$-ray emitting blazars are extracted, and from the surface density of 
the three radio surveys. Contrary to the radio surveys that reach an almost homogenous 
depth over their 
footprints, the limiting sensitivity of the WISE catalog in each band is not uniform on the sky 
(cp. with Figure~8 at the Explanatory Supplement to the AllWise Data Release 
Products\footnote{http://wise2.ipac.caltech.edu/docs/release/allwise/expsup/sec4$\_$2.html}).
An excerpt of the catalog of WIBRaLS is shown in Table~4. The catalog contains
the following columns: WIBRaLS unique name, AllWISE WISE name, Right 
Ascension and Declination of the WISE source position, values of the three WISE colors 
and their uncertainties (corrected for galactic absorption), values of the scores for the 
BZB, BZQ and Mixed regions of the {\it locus} model, WISE-based type and spectral class, 
name of the final radio counterpart associated to the WISE source and value of the
$q_{22}$ parameter. The whole catalog will be available through Vizier and as a Cone Search 
service through all tools compatible with the Virtual Observatory (VO) specifications.

\begin{table*}
	\begin{center}
	\resizebox{2\columnwidth}{!}{%
	\begin{threeparttable}
	\caption{Sample of rows of the catalog of WIBRaLS. The complete catalog in electronic
	format will be available on Vizier and as a Cone Search service through all VO-compatible tools.}
	\begin{tabular}{ccccccccccccccccc}
	\tableline
  	WIBRaLS\tnote{a}		&	WISE\tnote{b} 	& 	R.A.\tnote{c} 	 & Dec.\tnote{d} &  $c_{12}$\tnote{e}  & $\sigma_{c_{12}}$\tnote{f} 	 & 	$c_{23}$\tnote{g} 	 & 	
	$\sigma_{c_{23}}$\tnote{h}  & $c_{34}$\tnote{i} & $\sigma_{c_{34}}$\tnote{l}  & $s_{\mathrm{BZB}}$	\tnote{m}  & $s_{\mathrm{MIX}}$\tnote{n}  & 
	$s_{\mathrm{BZQ}}$	\tnote{o}  & Type\tnote{p}  & 
	Class\tnote{q}  & Radio\tnote{r}  & $q_{22}$\tnote{s} 	\\	
  	name 	& name	& 	deg 	 & deg &  mag  & mag 	 & 	mag	 & 	mag  & mag & mag  &  &   &   &   & & name  & 	\\	
	\tableline
  WB J0004-4736 & J000435.65-473619.6 & 1.149 & -47.605 & 1.09 & 0.03 & 3.32 & 0.03 & 2.31 & 0.07 & 0.0 & 0.0 & 0.83 & BZQ & B & SUMSSJ000435-473620 & -1.65\\
  WB J0005+1609 & J000559.23+160949.0 & 1.497 & 16.164 & 1.06 & 0.03 & 2.61 & 0.03 & 2.36 & 0.06 & 0.0 & 0.87 & 0.0 & MIXED & B & NVSSJ000559+160946 & -1.48\\
  WB J0005+3820 & J000557.18+382015.1 & 1.488 & 38.338 & 1.1 & 0.03 & 3.27 & 0.03 & 2.68 & 0.03 & 0.0 & 0.0 & 0.95 & BZQ & A & NVSSJ000557+382015 & -0.87\\
  WB J0005+5428 & J000504.36+542825.0 & 1.268 & 54.474 & 1.09 & 0.04 & 2.47 & 0.06 & 2.37 & 0.13 & 0.0 & 0.7 & 0.0 & MIXED & C & NVSSJ000504+542825 & -1.43\\
  WB J0005-1648 & J000517.92-164804.4 & 1.325 & -16.801 & 1.02 & 0.04 & 2.46 & 0.06 & 2.41 & 0.18 & 0.11 & 0.54 & 0.0 & MIXED & C & NVSSJ000517-164805 & -1.61\\
  WB J0005-2758 & J000558.54-275857.7 & 1.494 & -27.983 & 1.01 & 0.03 & 2.4 & 0.06 & 2.23 & 0.22 & 0.53 & 0.11 & 0.0 & BZB & C & NVSSJ000558-275900 & -1.79\\
  WB J0005-4518 & J000536.67-451845.6 & 1.403 & -45.313 & 1.11 & 0.05 & 2.69 & 0.12 & 2.64 & 0.33 & 0.0 & 0.08 & 0.32 & BZQ & C & SUMSSJ000537-451848 & -0.66\\
  WB J0005-6223 & J000527.13-622302.6 & 1.363 & -62.384 & 1.21 & 0.04 & 2.87 & 0.05 & 2.58 & 0.13 & 0.0 & 0.0 & 0.71 & BZQ & C & SUMSSJ000527-622302 & -0.73\\
  WB J0006+1235 & J000623.05+123553.1 & 1.596 & 12.598 & 1.29 & 0.03 & 2.81 & 0.03 & 2.32 & 0.07 & 0.0 & 0.0 & 0.41 & BZQ & C & NVSSJ000623+123553 & -1.03\\
  WB J0006+3422 & J000607.37+342220.4 & 1.531 & 34.372 & 1.06 & 0.04 & 2.66 & 0.06 & 2.03 & 0.21 & 0.32 & 0.21 & 0.0 & BZB & D & NVSSJ000607+342220 & -1.01\\
	\tableline
	\end{tabular}%
	\begin{tablenotes}[para]
 Ê Ê Ê Ê Ê	\item {Notes:}\\
 Ê Ê Ê Ê Ê	\item[a] WIBRaLS name (IAU format)\\
 Ê Ê Ê Ê Ê	\item[b] WISE name\\
		\item[c] Right Ascension\\
		\item[d] Declination\\
		\item[e] $[3.4]\!-\![4.6]$ WISE color (corrected for galactic extinction)\\
		\item[f] Uncertainty on the $[3.4]\!-\![4.6]$ WISE color\\
		\item[g] $[4.6]\!-\![12]$ WISE color (corrected for galactic extinction)\\
		\item[h] Uncertainty on the $[4.6]\!-\![12]$ WISE color\\	
		\item[i] $[12]\!-\![22]$ WISE color (corrected for galactic extinction)\\
		\item[l] Uncertainty on the $[12]\!-\![22]$ WISE color\\	
		\item[m] Score for the BZB region of the {\it locus}\\
		\item[n] Score for the Mixed region of the {\it locus}\\
		\item[o] Score for the BZQ region of the {\it locus}\\	
		\item[p] Spectral type (see Section~\ref{subsec:wise_selection})\\
		\item[q] Class (see Section~\ref{subsec:wise_selection})\\
		\item[r] Name of the radio counterpart\\
		\item[s] $q_{22}$ value\\
	\end{tablenotes}
	\end{threeparttable}	
	}
	\end{center}
	\label{tab:catalog}
\end{table*}

\section{Discussion}
\label{sec:discussion}

The nature of candidate blazars can only be confirmed through optical spectroscopic 
follow-up observations or by collecting extensive multi-wavelength photometric 
data to model their spectral energy distribution. Several recent efforts have 
validated the nature of a large
number of candidate blazars, selected with different techniques and associated to 
unidentified $\gamma$-ray sources in the 2FGL, using new and archival spectroscopic 
data~\citep{shaw2013,masetti2013,paggi2014,landoni2014,massaro2014b}. Nonetheless,
given the current lack of observations that 
would firmly establish the nature of the whole sample of WIBRaLS, the only 
valuable information about the nature of the catalog discussed in this paper can be indirectly inferred 
by comparison of our sample with similarly WISE-based AGN selection 
techniques (Section~\ref{subsec:comparison}). 
The comparison with the AGN 
selection techniques allows us to rule out significant contamination from non-AGNs, especially 
in the more numerous subclass of WIBRaLS classified as BZQ. This comparison also
underlines the significant
difference in the WISE colors of the WIBRaLS classified as BZB (BL Lacs) compared to the 
FSRQs subclass and the general population of radio-quiet mid-IR AGNs.
Finally, in order to identify the possible contamination in our WIBRaLS catalog from AGNs not
classified as blazars, in Section~\ref{subsec:veroncat} we also discuss the intersection of our 
sample and one of the largest compilation of AGNs, QSOs and BL Lacs available, 
the VERONCAT~\citep{veron2010}.

\subsection{Comparison with Other WISE-based AGNs Selection Techniques}
\label{subsec:comparison}

Selection techniques for AGNs based on their photometric mid-IR properties have become commonplace 
with the availability of WISE data. Most such techniques have been
fine-tuned to identify AGNs usually selected by other mid-IR colors ({\it Spitzer}) and their X-ray 
emission. In this Section, we will compare the sample of WIBRaLS with WISE-based 
selection criteria from~\cite{jarrett2011,stern2012,mateos2012,assef2013}.
In the following, the essential description of each of these selection techniques will be given.

\begin{itemize}
	\item[(A)]~\cite{jarrett2011} defined a region of the $[4.6]\!-\![12]$ vs $[3.4]\!-\![4.6]$ WISE 
	color-color diagram (the WISE AGNs ``box''), using the {\it Spitzer} colors classification criteria 
	determined by~\cite{stern2005} to validate the WISE selection. The AGNs box is defined as the 
	region of the $[4.6]\!-\![12]$ vs $[3.4]\!-\![4.6]$ diagram bounded by the constant lines: 
	$[4.6]\!-\![12]\!>\!2.2$, $[4.6]\!-\![12]\!<\!4.2$ and $[3.4]\!-\![4.6]\!<\!1.7$ and the relation 
	$[3.4]\!-\![4.6]\!>\!0.1([4.6]\!-\![12])\!+\!0.38$.
	\item[(B)]~\cite{stern2012} defined a WISE color criterion $[3.4]\!-\![4.6]\!\geq\!0.8$ for 
	sources with magnitude $[4.6]\!<\!15.0$, which identifies $\sim\!62\!\pm\!5.4$ AGNs per 
	deg$^{2}$. This condition selects $78\%$ of the {\it Spitzer} AGN candidates selected
	as discussed in~\cite{stern2005}, with an efficiency of $95\%$.
	\item[(C)]~\cite{assef2013} extended the selection proposed by~\cite{stern2012} based on the 
	WISE photometry in the $[3.4]$ and $[4.6]$ bands to provide a parametrized criterion that can 
	be changed to maximize either the efficiency or the completeness of the selection. The general 
	form of the constraint is $[3.4]\!-\![4.6]>\alpha_{R}\exp{[\beta_R([4.6]-\gamma_{R})^2]}$. In this 
	paper, we have used two different sets of the parameters $[\alpha_{R},\beta_{R},\gamma_{R}]$, 
	namely $[\alpha_{R},\beta_{R},\gamma_{R}]\!=\![0.662,0.232,13.97]$ and 
	$[\alpha_{R},\beta_{R},\gamma_{R}]\!=\![0.530,0.183,13.76]$, which yield efficiencies 
	90\% and 75\%, respectively~\citep{assef2013}. 
	\item[(D)]~\cite{mateos2012} presented a selection based on the WISE magnitudes in the 
	$[3.4]$, $[4.6]$ and $[12]$ bands, defined in a wedge bounded by the following constraints: 
	$[4.6]\!-\![12]\!\geq\!2.517$ and 
	$0.315([4.6]\!-\![12])\!-\!0.222\!<\![3.4]\!-\![4.6]\!<\!0.315([4.6]\!-\![12])\!+\!0.796$. This method
	takes into account uncertainties on the WISE photometry of the sources and is able to 
	recover $\sim97\%$ and $\sim77\%$ of the type-1 and type-2 bona-fide AGNs observed in the 
	Bright Ultra-Hard XMM-{\it Newton} Survey~\citep{mateos2013}. The same authors
	discuss a modified WISE-based AGN selection techniques that employs photometry in all four 
	WISE filters. We will not consider this further selection technique in this paper because it attains 
	smaller efficiency and completeness~\citep{mateos2012}.
\end{itemize}

\begin{figure}[]
	\includegraphics[height=8cm,width=8.5cm]{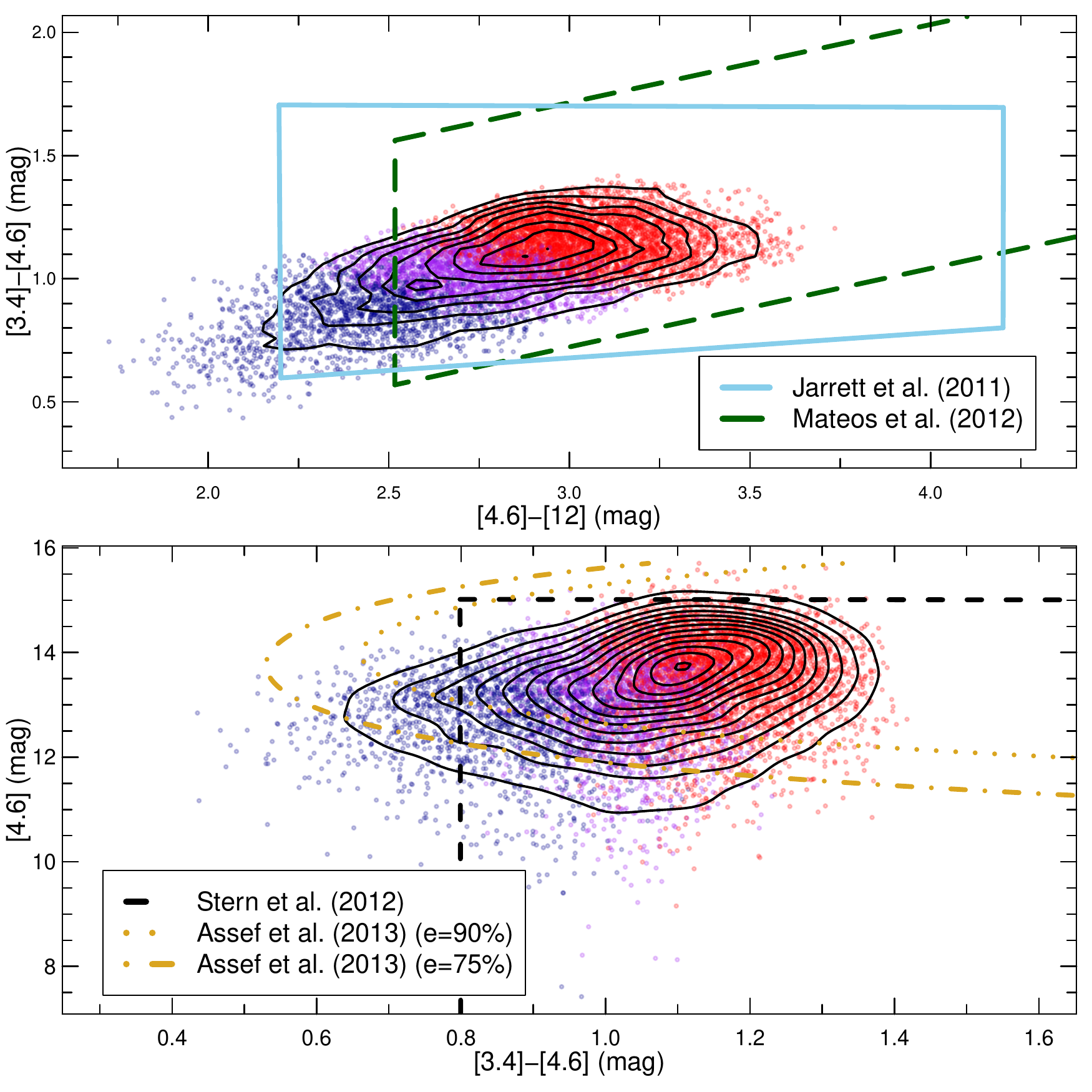}
	\caption{Upper panel: projection of the distribution of WIBRaLS onto the 
	$[4.6]\!-\![12]$ vs $[3.4]\!-\![4.6]$ WISE color-color plane. The regions used by the AGN selection
	techniques from~\cite{jarrett2011} and~\cite{mateos2012} are also shown. Lower panel: projection 
	of the distribution of WISE blazar-like radio sources onto the $[3.4]\!-\![4.6]$ vs $[4.6]$ WISE 
	color-magnitude plane, showing also the curves used to select AGNs according 
	to~\citep{stern2012} and~\cite{assef2013} (the yellow dotted line represent the selection 
	with efficiency $e\!=\!90\%$, the dashed-dotted line the selection with $e\!=\!75\%$). In both 
	panels, the WIBRaLS are color-coded according to their WISE-based spectral classification, 
	with red, bue and magenta symbols associated to sources classified as BZQ, BZB and 
	Mixed, respectively. Moreover, the two-dimensional density distributions of the whole WIBRaLS
	catalog in the two planes are represented by the isodensity contours
	(black solid lines).}
	\label{fig:colorcolormagcolor}
\end{figure}

The projections of the WIBRaLS distribution onto the WISE $[4.6]\!-\![12]$ 
vs $[3.4]\!-\![4.6]$ color-color plane and the $[3.4]\!-\![4.6]$ vs $[4.6]$ color-magnitude plane are shown 
in Figure~\ref{fig:colorcolormagcolor}. The regions of the two planes used to select the AGNs according to 
the four AGN selection methods described above are overplotted to the distribution of the WISE 
blazar-like radio sources. 

\begin{figure}[]
 	\includegraphics[height=8cm,width=8.5cm]{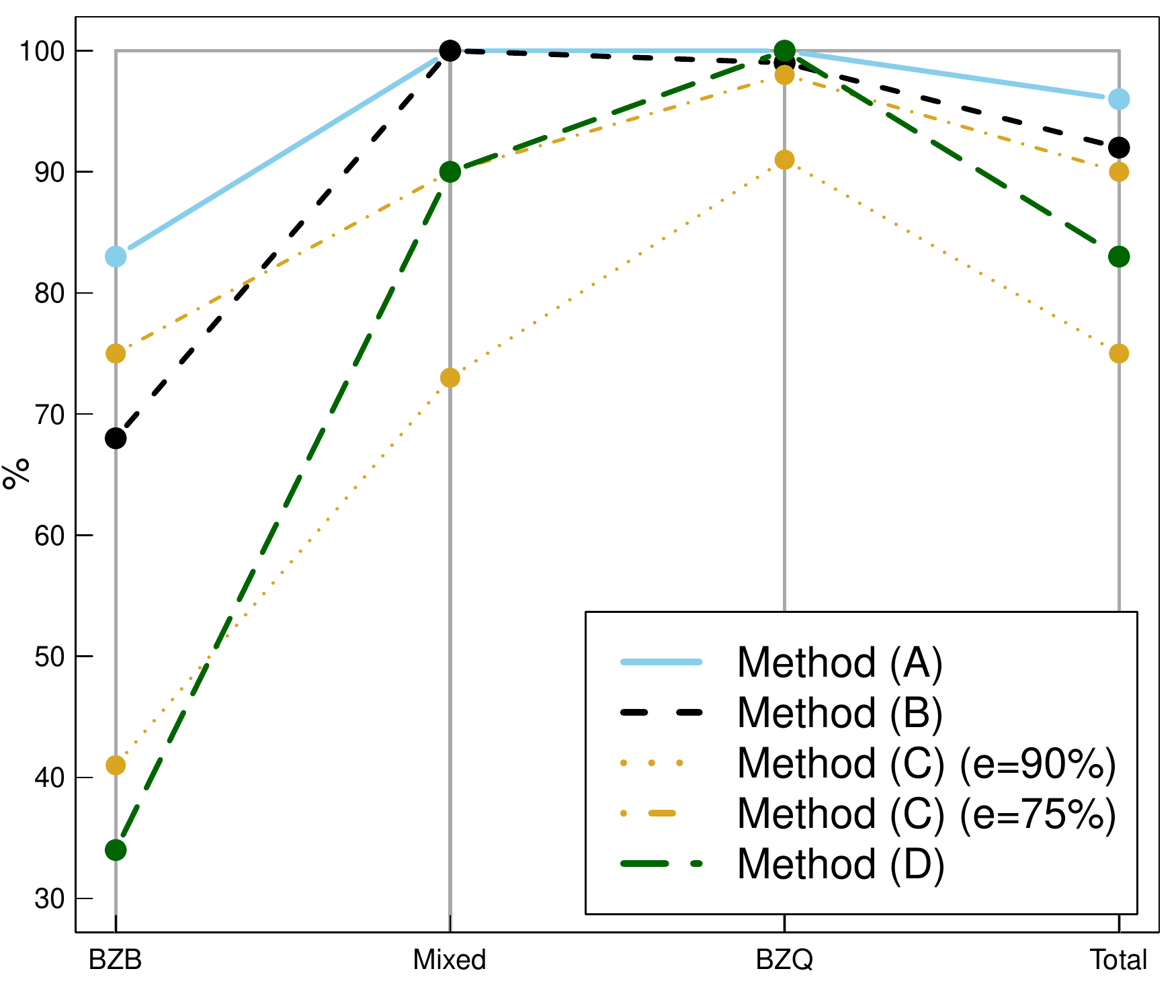}
	\caption{Fractions of WIBRaLS, split according to the spectral class and for the 
	whole sample, selected as AGNs by the five methods discussed in Section~\ref{subsec:comparison}.}
        \label{fig:quasars}
\end{figure}

We have applied the four distinct AGNs selection criteria to the 7855 members of the 
catalog of WIBRaLS. The numbers and fractions of WIBRaLS selected as candidate AGNs 
by each of the four AGN selection methods, split according to their spectral classification 
based on their WISE colors, are reported in Table~\ref{tab:comparison} and shown in 
Figure~\ref{fig:quasars}. 

The selection methods from~\cite{jarrett2011} and~\cite{stern2012} recover the largest fractions of 
WIBRaLS sources
($97\%$ and $94\%$ respectively). The criterion from~\cite{assef2013} with efficiency $e\!=\!75\%$, 
selects $90\%$ of the whole sample of WIBRaLS, while the same method with 
selection efficiency $e\!=\!90\%$ and the method from~\cite{mateos2012}, both recover $75\%$ and 
$85\%$ of our catalog, respectively. Figure~\ref{fig:quasars} 
shows the percentage of WIBRaLS selected as AGN candidates by each methods 
for each WISE spectral class. The fractions of WIBRaLS classified as BZQ and in the 
Mixed region that are selected as AGNs by either of the four methods discussed is larger than $90\%$, 
while the corresponding fractions of WIBRaLS classified as BZB are significantly lower, spanning from
$\sim\!35\%$ of the~\cite{mateos2012} method to the the $\sim\!84\%$ of the~\cite{jarrett2011} method.

The total fraction of WIBRaLS sources not selected by either one of these AGN selection techniques is 
$\sim\!5\%$, almost entirely ($\sim\!94\%$) classified as BZB by our method (Figure~\ref{fig:colorcolormagcolor}).
The total fraction of WIBRaLS classified as BZB selected by any of the five AGN selection techniques
used is $\sim\!68\%$. On the other hand, the fact that WIBRaLS classified as BZQ are 
overwhelmingly selected as AGNs by each of the methods considered (see Table~\ref{tab:comparison}) 
is explained by the large overlap between the region occupied by the WIBRaLS 
classified as candidate BZQ and the AGNs box in the two-dimensional WISE $[3.4]\!-\![4.6]$ vs $[4.6]\!-\![12]$ 
color space~\citep[see upper panel in Figure~\ref{fig:colorcolormagcolor} and discussion in][]{wright2010,jarrett2011,dabrusco2012}. The BZQ-type WIBRaLS 
have mid-IR properties which are similar to the mid-IR and optical properties of generic quasars and, more 
specifically, to the population of radio-quiet AGNs 
usually selected by mid-IR based techniques. The contamination from non-FSRQs in the WIBRaLS 
catalog is further reduced by selecting only sources that can be positionally associated to radio counterparts and 
with $q_{22}$ values compatible with the values of confirmed blazars. A relevant fraction of the WIBRaLS 
classified as BZB (from $\sim\!15\%$ to $\sim\!65\%$ for the AGN selection methods 
in Table~\ref{tab:comparison}) are bluer than the typical AGNs in WISE and, for this reason, are not 
selected by the techniques discussed in this Section.

\begin{table*}
	\begin{center}
	\caption{Fraction of WIBRaLS selected as AGNs by three distinct
	WISE-based selection methods: method (A)~\citep{jarrett2011}, method (B)~\citep{stern2012}, 
	method (C)~\citep{assef2013} and method (D)~\citep{mateos2012}. The selection technique 
	discussed in~\citep{assef2013} is applied with two different sets of parameters associated to 
	efficiencies $e\!=\!90\%$ and $e\!=\!75\%$ respectively. In parenthesis, the fraction of selected 
	sources relative to the number of WIBRaLS in each spectral class.}
	\begin{tabular}{lcccc}
	\tableline
	\tableline
						& \multicolumn{4}{c}{WISE}												  		\\
	\tableline
						& BZB				& Mixed				& BZQ	  			& 	Total			\\	
	\tableline
	Method (A)			&  1396 ($\sim 83\%$)	&  2194 ($100\%$)	& 3979 ($100\%$)		& 7569($\sim 96\%$)	\\	
	Method (B)			&  1136 ($\sim 68\%$)	&  2192 ($\sim 100\%$)&  3928 ($\sim99\%$)	& 7526($\sim 92\%$)	\\
	Method (C) $(e\!=\!90\%)$	& 682 ($\sim 41\%$)	&  1599 ($\sim 73\%$)	& 3612 ($\sim973\%$)	& 5893($\sim 75\%$)	\\
	Method (C) $(e\!=\!75\%)$	& 1238 ($\sim 74\%$)	&  1982 ($\sim 90\%$)	& 1982($\sim90\%$)		&  7100($\sim 90\%$)	\\	
	Method (D)			& 565 ($\sim 354\%$)	& 1978 ($\sim90\%$)	&  3979($\sim100\%$)		&  6522($\sim 83\%$)	\\	
	\tableline
	\label{tab:comparison}
	\end{tabular}
	\end{center}	
\end{table*}

\subsection{Cross-match with the VERONCAT}
\label{subsec:veroncat}

We have positionally cross-matched the WIBRaLS catalog with the Veron Catalog of 
Quasars \& AGN (VERONCAT)~\citep{veron2010} (v13). The VERONCAT is an inhomogeneous 
compilation of known AGNs, containing $\sim\!1.7\times\!10^{5}$ sources in its 13th version. 
Sources in the VERONCAT are broadly classified in AGNs (i.e., Seyfert galaxies and LINERs faint 
than $M_{B}=-22.25$), 
QSOs (star-like sources with absolute magnitude $M_{B}\!<\!-22.25$ and broad emission lines) 
and BL Lacs (confirmed or potential BL Lacs based on optical and radio observations). For the crossmatch, 
we used a conservative maximum radius of 1\arcsec around the position of the radio counterparts for 
each WIBRaLS, since all coordinates reported in the VERONCAT are based 
on radio or optical observations and their positional uncertainties are systematically smaller than 
1\arcsec.\footnote{Using a radius of 3.3\arcsec, we found 1618 unique crossmatches with 
VERONCAT sources. The fractions of different WISE spectral classes of the WIBRaLS 
associated with this larger radius, and the composition of the crossmatches in terms of the 
classification available in the VERONCAT are similar to the ones described for the smaller sample
obtained with radius 1\arcsec.} 
We found counterparts in VERONCAT for 797 WIBRaLS ($\sim\!10.1\%$ of the total 
number of WIBRaLS): based on the classification available in the VERONCAT, 
$57\%$ of the counterparts are classified as quasars, $13\%$ as Seyferts, $14\%$ as 
BL Lacs and the remaining $\sim\!16\%$ is classified as generic AGN. It is relevant to emphasize that
the classification available in the VERONCAT are not always reliable. For example, a non-negligible
fraction of sources classified as BL Lacs in the VERONCAT are not classified as blazars in the 
ROMA-BZCat~\citep{massaro2011}. In terms of the WISE 
spectral classification, the cross-matched WIBRaLS 
are split in 512 BZQ ($\sim\!52\%$) candidates, 213 BZB ($\sim\!20\%$) candidate blazars and 
the remaining 70 WIBRaLS are located in the Mixed region of the {\it locus} model. The fact that the 
$\sim\!90\%$ of WIBRaLS cannot be associated to a VERONCAT source
likely depends on the lack of of optical spectroscopic observations and redshift measurements that are 
required for a source to be included in the VERONCAT. In particular, given that the optical spectra of 
BL Lacs are typically featureless and the estimation of the redshift can be problematic, we expect the 
VERONCAT to miss a large fraction of confirmed blazars. This conclusion is further reinforced by 
noticing that the positional crossmatch between the most recents version of the ROMA-BZCat and 
VERONCAT within 1\arcsec, only returns 1649 crossmatches on 3149 total sources, corresponding to 
$\sim\!52.4\%$ of the ROMA-BZCat. 

Using the optical B magnitude and radio flux density at 6cm available for a subset of sources in 
the VERONCAT, we defined as radio-loud AGNs the sources with radio-loudness parameter
$\log{R}\!\geq\!1$ where $R\!=\!S_{\mathrm{6cm}}/S_{\mathrm{B}}$ is the ratio between the 6cm radio
and the B flux densities. We assumed that sources with no radio flux measurement are radio-quiet.
We found that 644 out of the 797 total crossmatches ($\sim\!81\%$) between the WIBRaLS catalog and 
VERONCAT obtained with maximum radial distance 1\arcsec\ are radio-loud, suggesting
a contamination from radio-quiet AGNs in our catalog smaller than $\sim\!19\%$. We also checked
the contamination from SSRQs in our sample by determining the radio spectral index
$\alpha_{R}$ between the flux densities measured at 20cm and 6cm, both available for all 644 
VERONCAT-WIBRaLS counterparts. We have considered SSRQs the sources with $\alpha_{R}\!>\!0.5$. 
We found only 73 source which can be classified as SSRQs, corresponding to $\sim\!11\%$ of the 
total number of VERONCAT counterparts. A more detailed quantitative estimate of the contamination 
in the WIBRaLS sample introduced by SSRQs is made difficult by the shortage of large
catalogs of SSRQs. Nonetheless, based on the result obtained from the VERONCAT, we can assume 
that the fraction of SSRQs in the WIBRaLS sample will not exceed $\sim\!10\%$. Furthermore, 
it has been shown that confirmed BL Lacs from the ROMA-BZCat can have steep radio 
spectra~\citep[see, for instance, Figure 11 in][]{massaro2013d}, making the observational 
differences in the definition of the two classes of AGNs less sharp. This fact, in principle, 
indicates that the effective contamination from SSRQs is smaller than the nominal $\sim\!11\%$ 
determined here.

We do not observe significant differences between the distributions of the WISE colors and fluxes 
of the sources that have been associated to a VERONCAT counterparts and the remainder of the 
WISE blazar-like radio sources. As expected, the WIBRaLS classified as BZB-type 
candidate blazars are overwhelmingly ($\sim92\%$) associated with VERONCAT counterparts classified 
as BL Lac, while the fractions of VERONCAT counterparts classified as BL Lac and associated to WIBRaLS 
of BZQ-type or Mixed are small ($\sim\!6\%$ and 
$\sim\!19\%$, respectively). Even though our results are based on a small
subset of the catalog of WIBRaLS crossmatched with a VERONCAT source, 
they indicate that the fraction of AGNs not classified as blazars is much larger for BZQ and Mixed 
candidate blazars than for BZB candidate blazars.

\section{Summary and Conclusions}
\label{sec:summary}

We have presented a catalog of candidate $\gamma$-ray blazars extracted from the 
AllWISE WISE Data Release, using a modified version of the association method for $\gamma$-ray
unidentified sources from~\cite{dabrusco2013}. This method is based on a model of the three-dimensional
{\it locus} occupied by the confirmed $\gamma$-ray emitting blazars in the WISE color space. The 
WISE blazar-like sources have been spatially associated to radio sources in either the NVSS, 
FIRST or SUMSS surveys. A further selection based on their radio-loudness has been performed using 
the $q_{22}$ parameter to retain only radio-loud AGNs in our sample, 
and minimize the contamination from other extragalactic radio sources. The main results of this paper
can be summarized as follows:

\begin{itemize}
	\item The final catalog of unique WIBRaLS contains 7855 sources, split in 1682 BZB-type 
	candidate blazars, 3973 BZQ-type candidate blazars and 2194 candidate blazars classified 
	as Mixed. The sky distribution of the members of the catalog reflects
	the coverage of the radio surveys used and the variable limiting sensitivity of the WISE survey, 
	especially in the $[22]$ filter. 
	\item Out of the total 7855 WIBRaLS, 1295 sources $\sim\!16.5\%$ can be spatially 
	associated to {\it bona-fide} blazars in the ROMA-BZCat. The number of WIBRaLS that 
	can be crossmatched with one confirmed $\gamma$-ray 
	emitting blazars extracted from the ROMA-BZCat and used to define the model of the {\it locus} is 454
	($\sim76.5\%$ of the {\it locus} sample). Moreover, 797 WIBRaLS 
	(mostly classified as BZQ according to our method based on WISE colors) can be 
	cross-matched to VERONCAT sources within 1\arcsec. Their VERONCAT counterparts are 
	classified as quasars ($57\%$), Seyferts ($13\%$), BL Lacs ($14\%$) and
	generic AGN ($\sim\!16\%$).
	\item The comparison of the catalog of WIBRaLS with other WISE-based
	AGN selection techniques suggests that the contamination from non-AGNs in our catalog is 
	very low. While almost all the WISE blazar-like sources classified as BZQ (representing $\sim50\%$
	of the total sample) are selected by either one of the other techniques, the fraction of WIBRaLS 
	classified as BZB that are recovered by the other methods is significantly lower 
	(from $\sim\!30\%$ to $\sim\!70\%$). These difference depends on the fact that BL Lacs 
	candidates (BZB-type WIBRaLS) are 
	significantly bluer than the typical AGNs in WISE, especially in the $[3.4]$ and $[4.6]\mu$m 
	filters.
	\item We have estimated the contamination in the WIBRaLS catalog from radio-quiet 
	AGNs and SSRQs. Using a sample of 797 WIBRaLS spatially associated to
	VERONCAT sources, we found that only $\sim\!19\%$ of our sample can be considered
	radio-quiet. This fraction represents an upper limit to the contamination because we considered
	VERONCAT sources missing radio flux density measurements at 6cm radio-quiet. We also
	confirmed that SSRQs can contaminate the WIBRaLS catalog using a {\it bona find} sample of 
	18 SSRQs produced by~\cite{gu2013}. Also in this case, using the VERONCAT, we showed 
	that the fraction of SSRQ contaminants in our catalog does not exceed $\sim\!10\%$. 	
\end{itemize}

Our catalog of WISE-selected $\gamma$-ray blazar candidates is intended to be a useful 
resource for the future investigations of the unidentified sources detected in both the $\gamma$-ray 
and X-ray energies. In this paper we do not address the problem of quantifying the effect of 
several factors that can, in principle, prevent a $\gamma$-ray blazar candidate from our catalog to 
be observed in the current and future $\gamma$-ray observations (including variability, intrinsic 
scatter in the distribution of mid-IR vs $\gamma$-ray fluxes for confirmed blazars and the procedure 
for the source detection in the {\it Fermi} LAT data). We expect nonetheless that the catalog 
of WIBRaLS will provide astronomers with a valuable resource to constrain the nature of 
unidentified high-energy sources.

As an example,~\cite{paggi2013} have shown that a large fraction of the 
unassociated $\gamma$-ray sources located in regions of the sky observed by the X-Ray 
Telescope (XRT) on board of the {\it Swift} satellite, can be associated to X-ray counterparts 
whose coordinates are compatible with the position of WISE candidate blazars selected
with the method discussed by~\cite{dabrusco2013}. Moreover,~\cite{maselli2013} have used WISE 
candidate $\gamma$-ray emitting blazars to associate 24 unidentified hard X-ray sources of the 
Third Palermo {\it Swift} Burst Alert Telescope 
(BAT)\footnote{http://www.ifc.inaf.it/cgi-bin/INAF/pub.cgi?href=activities/bat/index.html}.~\cite{cowperthwaite2013} have also used
the method for the extraction of candidate $\gamma$-ray emitting blazars based on the {\it locus} 
in the WISE colors space to associate 13 sources, extracted from the Astronomer's Telegrams, 
that exhibit non-periodic variability, mostly at high-energy. These examples clearly show that 
WISE-based selection of $\gamma$-ray emitting candidate blazars can be successfully 
used to constrain the nature of unidentified high-energy sources observed in different spectral 
ranges and with different techniques.

\acknowledgements

The authors thank the anonymous referee for the insightful comments that have helped to 
significantly improve the manuscript. 
This investigation is supported by the NASA grants NNX12AO97G and NNX13AP20G. The work by G. 
Tosti is supported by the ASI/INAF contract I/005/12/0. H. A. Smith acknowledges partial support from 
NASA/JPL grant RSA 1369566. This research has made use of data obtained from the high-energy 
Astrophysics Science Archive Research Center (HEASARC) provided by NASA's Goddard Space Flight 
Center; the SIMBAD database operated at CDS,
Strasbourg, France; the NASA/IPAC Extragalactic Database
(NED) operated by the Jet Propulsion Laboratory, California Institute of Technology, under contract 
with the National Aeronautics and Space Administration. This publication makes use of data products 
from the Wide-field Infrared Survey Explorer, 
which is a joint project of the University of California, Los Angeles, and 
the Jet Propulsion Laboratory/California Institute of Technology, 
funded by the National Aeronautics and Space Administration.
Part of this work is based on the NVSS 
(NRAO VLA Sky Survey): The National Radio Astronomy Observatory is operated by Associated Universities,
Inc., under contract with the National Science Foundation and on the VLA low-frequency Sky Survey (VLSS).
The Molonglo Observatory site manager, Duncan Campbell-Wilson, and the staff, Jeff Webb, 
Michael White and John Barry, are responsible for the smooth operation of Molonglo Observatory 
Synthesis Telescope (MOST) and the day-to-day observing programme of SUMSS. 
The WENSS project was a collaboration between the Netherlands Foundation 
for Research in Astronomy and the Leiden Observatory. 
We acknowledge the WENSS team consisted of Ger de Bruyn, Yuan Tang, 
Roeland Rengelink, George Miley, Huub Rottgering, Malcolm Bremer, 
Martin Bremer, Wim Brouw, Ernst Raimond and David Fullagar 
for the extensive work aimed at producing the WENSS catalog.
TOPCAT\footnote{\underline{http://www.star.bris.ac.uk/$\sim$mbt/topcat/}} 
\citep{taylor2005} for the preparation and manipulation of the tabular data and the images.

{}

\end{document}